\documentclass[aps,pre,superscriptaddress,twocolumn,floatfix]{revtex4-2}
\usepackage[margin=1in]{geometry}
\usepackage{amsmath,amsfonts, setspace,graphicx,comment,chemformula}
\usepackage{soul,xcolor,algpseudocode,algorithm,bm}
\usepackage{multirow,hhline}
\usepackage[section]{placeins}

\DeclareMathOperator*{\argmin}{arg\,min}
\newcommand{\be}{\begin{equation}}
\newcommand{\ee}{\end{equation}}
\newcommand\norm[1]{\lVert#1\rVert}
\newcommand{\lbc}[1]{{\color{red} [LB: #1]}}   
 
\begin{document}

\title{Exploring Bayesian olfactory search in realistic turbulent flows}
\author{R.\ A.\ Heinonen}
\affiliation{Dept.\ Physics and INFN, University of Rome ``Tor Vergata'', Via della Ricerca Scientifica 1, 00133 Rome, Italy}
\author{L.\ Biferale}
\affiliation{Dept.\ Physics and INFN, University of Rome ``Tor Vergata'', Via della Ricerca Scientifica 1, 00133 Rome, Italy}
\author{A.\ Celani}
\affiliation{Quantitative Life Sciences, The Abdus Salam International Centre for Theoretical Physics, 34151 Trieste, Italy}
\author{M. Vergassola}
\affiliation{Laboratoire de physique, \'Ecole Normale Sup\'erieure, CNRS, PSL Research University, Sorbonne University, Paris 75005, France}
\begin{abstract}
The problem of tracking the source of a passive scalar in a turbulent flow is relevant to flying insect behavior and several other applications. Extensive previous work has shown that certain Bayesian strategies, such as ``infotaxis,'' can be very effective for this difficult ``olfactory search'' problem. More recently, a quasi-optimal Bayesian strategy was computed under the assumption that encounters with the scalar are independent. However, the Bayesian approach has not been adequately studied in realistic flows which exhibit spatiotemporal correlations. In this work, we perform direct numerical simulations (DNS) of an incompressible flow at $\mathrm{Re}_\lambda\simeq150,$ while tracking Lagrangian particles which are emitted by a point source and imposing a uniform mean flow with several magnitudes (including zero). We extract the spatially-dependent statistics of encounters with the particles, which we use to build Bayesian policies, including generalized (``space-aware'') infotactic heuristics and quasi-optimal policies.  We then assess the relative performance of these policies when they are used to search using scalar cue data from the DNS, and in particular study how this performance depends on correlations between encounters. Among other results, we find that quasi-optimal strategies continue to outperform heuristics in the presence of strong mean flow but fail to do so in the absence of a mean flow. We also explore how to choose optimal search parameters, including the frequency and threshold concentration of observation.
\end{abstract}
\maketitle

\section{Introduction}

\begin{figure}
    \centering
    \includegraphics[width=\linewidth]{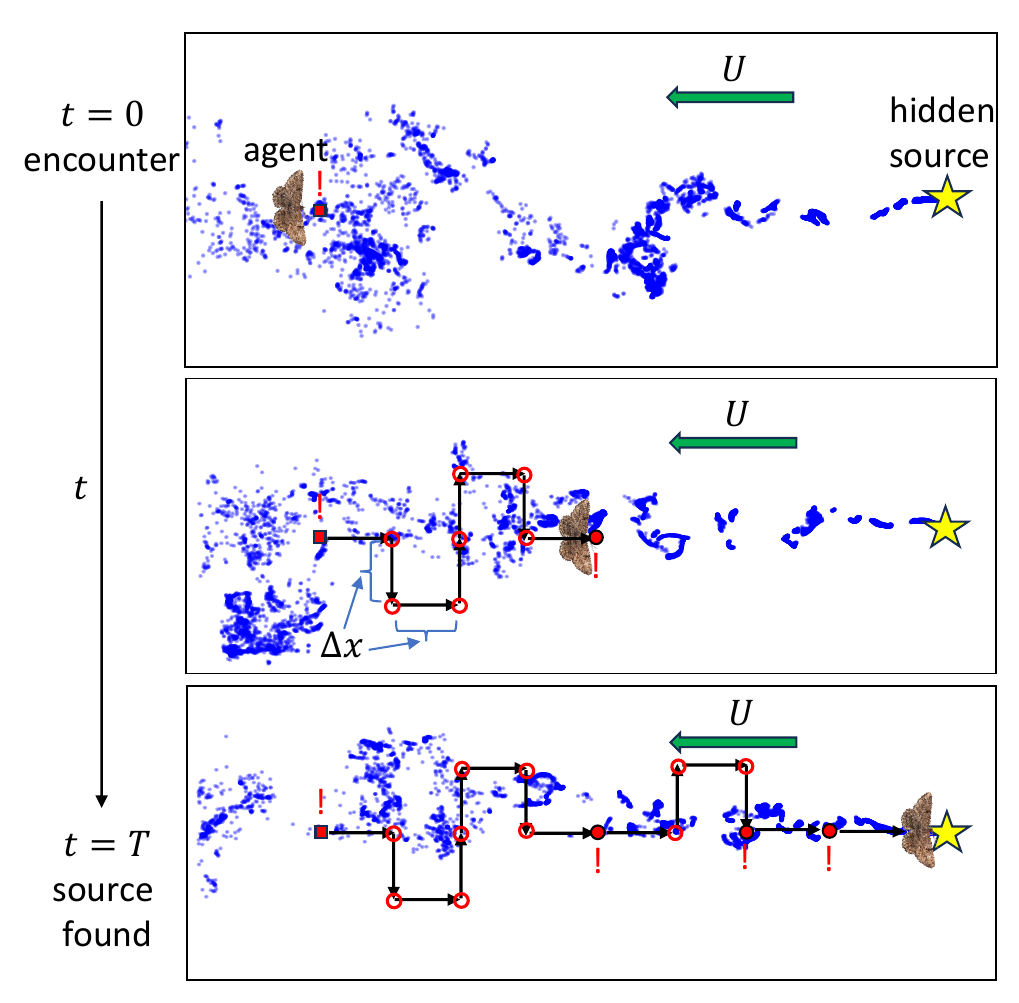}
    \caption{Cartoon illustrating the search problem. An agent, symbolized by a moth, is sensitive to a passive scalar (blue dots) being advected by a turbulent flow and must decide how to move in order to arrive at the unseen, immobile source of the passive scalar (yellow star) in minimal time. In our model-based formulation, the agent only makes binary (low/high) measurements of the local concentration (unfilled/filled red circles) but has access to the likelihood of these measurements conditioned on the agent's position relative to the source. The search begins with an encounter (high measurement) at time zero, and the agent moves a fixed distance $\Delta x=v_A \Delta t$ every timestep $\Delta t,$ where $v_A$ is the agent speed. Unless otherwise stated, the measurements are also taken at every $\Delta t.$ The concentration snapshots are taken from our simulations at $U=4.9 u_{\rm rms}.$ }
    \label{fig:search_cartoon}
\end{figure}
Suppose an agent in a flow is sensitive to a passive scalar, say a chemical odor, which is emitted by a stationary, unseen point source (see Fig.~\ref{fig:search_cartoon}). When the flow is turbulent, finding the source using local concentration measurements is a notoriously difficult task \cite{reddy2022}. At sufficiently large distances from the source, encounters tend to be sparse and effectively randomized due to the stochastic and intermittent nature of passive scalar dispersion by turbulence  \cite{yee1993,celani2014}. Moreover, local concentration gradients do not generally point to the source instantaneously, and the mean gradients take too much time to converge to be exploitable \cite{bossert1963,murlis1981}. Searches over such long distances are relevant to a number of applications, including the mating and foraging behaviors of flying insects \cite{murlis1992} and aquatic animals \cite{michaelis2020}, as well as the robotic tracking of explosives \cite{nguyen2015} and hazardous contaminant leaks \cite{wiedemann2019}. 

In the literature, approaches to this source-tracking problem (often called ``olfactory search'') may be divided into two general categories: model-based methods, which assume access to a (possibly imperfect) model of the encounter statistics, and model-free methods, which relax this assumption. The former, which include the well-known and high-performing heuristic ``infotaxis'' \cite{infotaxis}, are typically Bayesian: a probability distribution (the ``posterior'' or ``belief'') over the possible source locations is maintained and updated using incoming observation data via Bayes' theorem. Recent progress \cite{loisy2022,heinonen2023,loisy2023} on the Bayesian approach has demonstrated that (quasi-)\textit{optimal} strategies (in the sense of minimal mean arrival time to the source) may be obtained using the formalism of partially observable Markov decision processes (POMDP) via approximate solution of the underlying Bellman equation \cite{sondik1978,shani_review}. Consistent with the claim of quasi-optimality, the computed policy outperforms all known heuristics, including both infotaxis and its even more performant variant, ``space-aware'' infotaxis (SAI) \cite{loisy2022}. 

However, in most studies on the model-based approach, the authors impose an analytic model for the likelihood of an encounter and draw encounters artificially from this model, rather than test search performance on realistic data from experiment or simulation. This approach has at least two drawbacks: for one, the models employed are typically very crude descriptions of turbulence, and model parameters are often drawn on seemingly arbitrary grounds; we have addressed this issue in a separate paper which studies quasi-optimal trajectories under realistic (single-time) flow statistics \cite{heinonen2024}. Secondly, and perhaps more seriously, even if one were assured a given model adequately depicted the statistics of a real flow, it would necessarily still miss the key detail that turbulent flows, and in particular the statistics of passive scalars \cite{shraiman_review}, exhibit spatiotemporal correlations, rendering the probability of an encounter dependent on the previous encounter history. In contrast, the Bayesian approach explicitly assumes encounters are independent, and its efficacy in the face of realistic correlations remains an important question. 

To be concrete, we will assume throughout this paper that the agent is searching for the source location in a turbulent, statistically stationary environment and registers an encounter if the local concentration of the passive scalar is instantaneously measured to be above some threshold $c_{\rm thr}$, so that the agent makes measurements of a signal $\Omega(\bm{x},t) = \theta(c(\bm{x},t) - c_{\rm thr})$ taking values in $\{0,1\}$ (here $\theta$ is the Heaviside function). 

We will denote the single-time likelihood, $ p_{\Omega}(\bm{x})$ at any position $\bm{x}$ with:
$$
\begin{cases}
    p_{1}(\bm{x}) = {\rm Pr}(\Omega=1) \\
    p_{0}(\bm{x}) = {\rm Pr}(\Omega=0), 
\end{cases}  
$$
with $p_0(\bm{x}) = 1-p_1(\bm{x})$. By ``single-time'' we mean that the probability to make an observation is given by the average signal in that position and is not conditioned on any observation data from previous times. We also assume the agent makes the observations every $\Delta t,$ or at a frequency $\nu_\Omega=\Delta t^{-1}.$ Letting $t_k = k \Delta t,$ the agent's observation sequence is then $\Omega_k = \Omega (\bm{x}(t_k),t_k),$ where $\bm{x}(t)$ is the agent's trajectory. In addition to affording simplicity and analytical tractability, the assumption of binary observations is plausible in realistic contexts in light of results using computational information theory which suggest that a very coarse measurement of the concentration field makes optimal use of the information present in the signal \cite{boie2018,victor2019}. We assume the agent always begins its search with an encounter, i.e., $\Omega_0 = 1.$ 

The agent maintains a spatial map $b(\bm{x})$ called the posterior or belief, which represents the probability density of the agent's position relative to the source, as inferred from the sequence of observations. It updates $b(\bm{x})$ after each observation $\Omega$ using Bayes' theorem:
\be \label{eq:bayes}
b(\bm{x} | \Omega) = \frac{b(\bm{x}) p_\Omega (\bm{x}) }{\int d{\bm x'} \, b(\bm{x}') p_\Omega (\bm{x'})}.
\ee
The likelihood of observation $p_\Omega(\bm{x})$ is therefore assumed to be known by the agent. As observation data arrive, one iteratively applies Eq.~\ref{eq:bayes}, and the posterior gradually sharpens and concentrates probability on the ground truth source location. However, this approach implicitly assumes that the observations are uncorrelated: in general, when inferring variable $y$ using observation data $z_1,\dots,z_n,$ Bayes' theorem gives
\begin{widetext}
\begin{equation}
{\rm Pr}(y|z_1,\dots,z_n) = \frac{{\rm Pr}(z_1,\dots ,z_n|y) {\rm Pr}(y)}{\int dy' \, {\rm Pr}(z_1,\dots ,z_n|y') {\rm Pr}(y')} \ne \frac{{\rm Pr}(z_1|y) \cdots {\rm Pr}(z_n|y)  {\rm Pr}(y)}{\int dy' \, {\rm Pr}(z_1|y') \cdots {\rm Pr}(z_n|y') {\rm Pr}(y')}
\end{equation}
\end{widetext}
unless the $z_i$ are conditionally independent, given $y$. This is manifestly false when the observation data are concentration measurements in a turbulent flow, which tends to organize passive scalars into coherent structures (see Fig.~\ref{fig:search_cartoon} for a visual aid).

The presence of correlations is intimately tied to the agent's search parameters---its speed, its observation frequency, and the threshold concentration to which it is sensitive---and the question of how these should be selected. Suppose the agent moves at fixed speed $v_A$ and makes observations at fixed rate $\nu_\Omega$. Moving with at higher speed (relative to that of the mean flow $\bm{U}$) will tend to decorrelate successive observations, whereas making observations at a higher rate will tend to correlate them more strongly. More precisely, one may define dimensionless parameters 
\be
\label{eq:dimensionless} \bar \nu = \nu_\Omega \tau_c; \;\bar v = \frac{|\bm{v}_A-\bm{U}|}{\nu_\Omega \ell_c},
\ee
where $\tau_c$ and $\ell_c$ are respectively a characteristic correlation time and correlation length associated with the concentration field; one expects correlations will have an impact if $\bar \nu \gtrsim 1$ and $\bar v \lesssim 1.$ Note that $\bar{v}$ depends on the direction of motion when there is a mean flow, so that the strength of correlations depends on the actions taken by the agent. 

The strength of correlations will also depend on the observation threshold $c_{\rm thr}.$ For instance, consider a simple advection-diffusion model of the transport
\be \label{eq:advectiondiffusion}
\partial_t c + \bm{U} \cdot \nabla c = D \nabla^2 c + S \delta^3(\bm{x}),
\ee
with $D$ an effective turbulent diffusivity and $S$ the emission rate of the source. Increasing the threshold as $c_{\rm thr} \to \lambda c_{\rm thr}$ can then be compensated by the transformation \begin{align*}c \to \lambda c; \, \bm{x} \to \lambda^{-1} \bm{x}; \, t \to \lambda^{-2} t; \, \bm{U} \to \lambda \bm{U},\end{align*} i.e., by decreasing length- and time-scales (including the diffusion length $D/U \to \lambda^{-1} D/U$). In particular, correlation lengths and times will shrink, weakening the effect of correlations. Although turbulent transport is more complex than diffusion, this basic qualitative principle will still hold in realistic flows.



It is clear then that one cannot hope to understand the impact of the choice of $v_A,$ $\nu_\Omega,$ and $c_{\rm thr}$ without understanding the impact of correlations on the search. In addition, the effect of correlations on the performance of the agent should be quantified. Thus, contact with realistic, correlated data is sorely needed. To this end, we performed high-fidelity direct numerical simulations (DNS) of a turbulent flow while tracking the positions of millions of Lagrangian tracer particles that are emitted from a stationary point source. To enable the study of searching within different plume structures, we impose a uniform mean flow with five different magnitudes (including the isotropic case without a mean flow). [These data were also used in Refs.~\cite{heinonen2024,piro2024}.]

\begin{figure*}
    \centering
    \includegraphics[width=\linewidth]{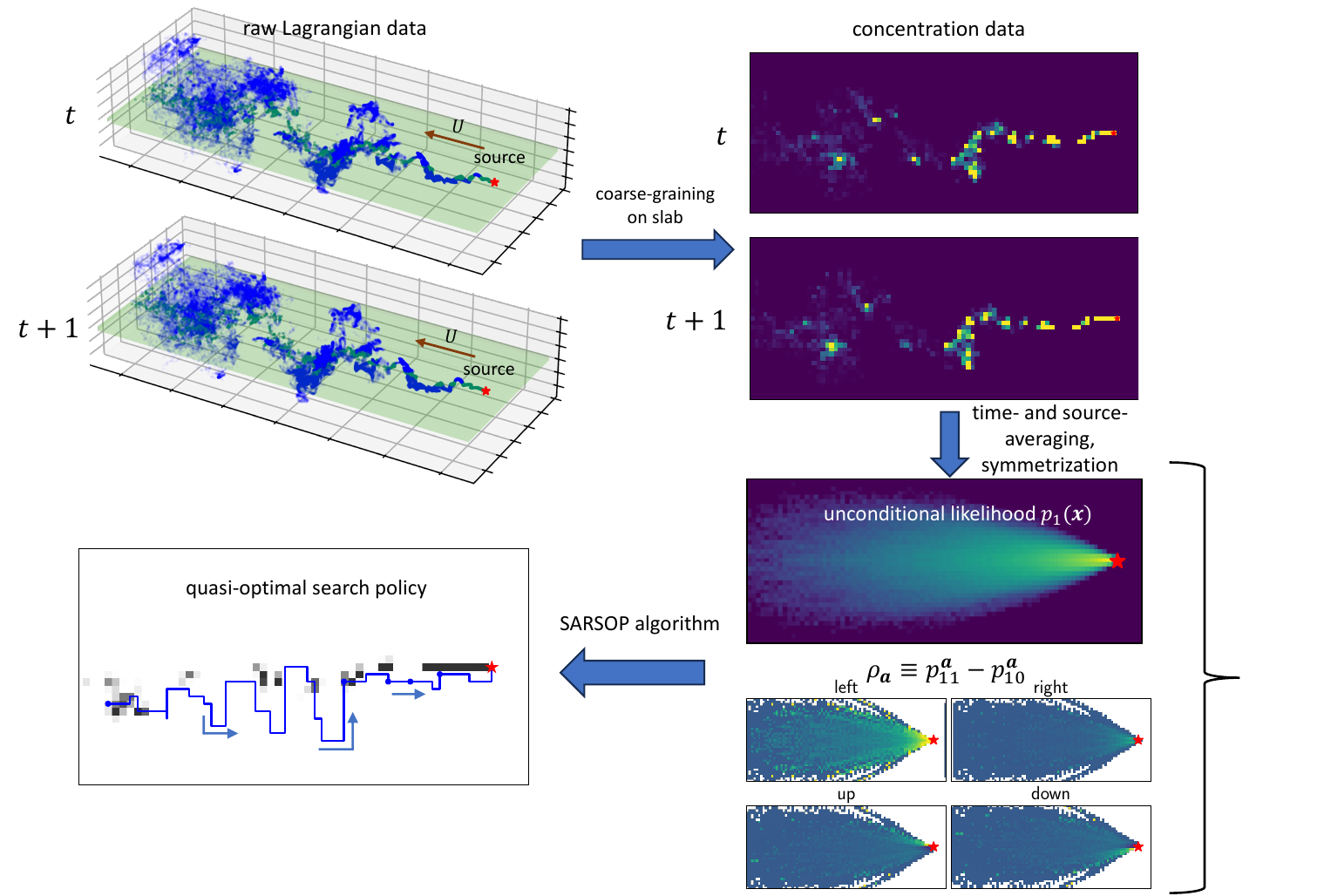}
    \caption{Schematic illustrating how the DNS data are used. First, the 3--D Lagrangian data are coarse-grained on a thin slab gridworld which is parallel to the mean wind $U$ and contains the source, thus obtaining a 2--D concentration field time series $c(\bm{x},t)$. The time series is averaged in time and symmetrized in space in order to obtain the space-dependent likelihood of the concentration being above a chosen threshold ${\rm Pr}(c(\bm{x})>c_{\rm thr})$ (as well as to compute an action-dependent correlation strength $\rho_{\bm{a}}$). The single-time likelihood (or in the case of correlation-aware policies, the two-time likelihood, obtainable from the correlation strength) is then used to compute a quasi-optimal policy using the SARSOP algorithm \cite{sarsop}, which yields an approximate solution to the Bellman equation. It is also used to perform Bayesian updates, for both the SARSOP policy and infotactic heuristics. Finally, the policies are used to search for the source, using the concentration time series to generate observations.}
    \label{fig:schematic}
\end{figure*}

We then pass to a POMDP setting by coarse-graining the particle data on a quasi-two-dimensional gridworld with thousands of cells. We define a threshold $c_{\rm thr}$ and obtain the empirical likelihood of an encounter, conditioned on the agent's position. This is used to build Bayesian policies; in this work, we study only the most performant: (space-aware) infotactic heuristics and quasi-optimal policies obtained with the SARSOP algorithm for solving POMDPs \cite{sarsop}. For completeness, we consider SAI to be a one-parameter family of policies and benchmark over a full scan of the parameter space. The complete pipeline is illustrated in Fig.~\ref{fig:schematic}.

We then compare how the policies perform when searching in the DNS data. First, we assess their relative quantitative performance in terms of the arrival time to the source. We pay particular attention to the effects of correlations, which we isolate by tuning their strength in two ways which do not change the likelihood of an encounter.

We summarize the key results as follows: 

(i) The dominant effect of the enhanced correlations is to increase the mean arrival time to the source. This effect is associated with small-scale flow structures and can be mostly explained by a simple first-order Markovian model of the correlations. 
An important exception to this rule occurs in the presence of a sufficiently strong mean flow: here, depending on the clock speed of the agent relative to that of the flow, we actually find that the mean arrival time can be substantially reduced relative to the uncorrelated baseline; in other words, the presence of correlations actually helps the agent find the source. This helpful effect is apparently associated with large-scale flow structures.

(ii) The relative performance of the quasi-optimal and infotactic policies is assessed, under the assumption of an agent unaware of correlations, i.e.,\ the model for the Bayesian inference is wrong. It is important to stress that there is no reason to expect \emph{a priori} that the policy obtained through SARSOP should outperform heuristics: the policy is optimized with respect to the \emph{uncorrelated problem} and has no performance guarantees in the presence of realistic data. However, we find that, in the presence of a sufficiently strong mean flow, the quasi-optimal policies continue to reliably outperform the entire SAI family, demonstrating an unexpected robustness. However, the performance under isotropic, zero-mean-flow conditions is qualitatively different. Here, SAI outperforms SARSOP by a considerable margin, and frequently pure infotaxis is the top performer in the SAI family. This is evidently a consequence of the presence of very strong correlations in the latter case, making the model of the environment used for the optimization too far from reality.

(iii) We address the natural question of whether strategies can be improved if the agent has some partial knowledge of the correlations and takes advantage of this knowledge during Bayesian inference and action selection. We speak about {\it correlation-aware agents} in this case. We show how to extend the POMDP approach to include temporal correlations up to first order at the Markov level and assess the relative improvement with respect to the performance of the {\it correlation-unaware agent.} 

(iv) Finally, we turn to the question of selecting agent parameters. We find evidence that, given a choice of agent speed $v_A$, correlations between observations induce an optimal observation rate $\nu_\Omega$; using search trials, we confirm that increasing the observation rate beyond the rate at which actions are taken does not improve and indeed even degrades search performance. Moreover, once $\nu_\Omega$ and $v_A$ have been fixed, we show that there is an optimal choice of the threshold $c_{\rm thr}.$

The remainder of the paper is organized as follows: in Sec.~\ref{app:theory}, we present calculations of the effect of correlations in the first-order Markov case, in order to help us understand how unmodeled correlations impact Bayesian search. In Sec.~\ref{sec:methods}, we present details on the DNS, how they were used to construct a POMDP, and how this POMDP was solved. In Sec.~\ref{sec:results} we first present arrival time statistics of search trials for different choices of policy and parameters, showing empirically how correlations impact the search when $\alpha$ is varied. We then discuss the parameter selection problem and support our conclusions with additional numerical experiments. Finally, in Sect.~\ref{sec:discussion} we summarize our findings and discuss future research avenues.

\section{Analysis: the impact of Markovian correlations on Bayesian source identification}\label{sec:theory}
To quantitatively understand how correlations impact search performance, it is useful to consider first the simplified set-up of a stationary agent. Let us assume that an agent located at position $\bm{x}^*$ relative to the source makes a discrete sequence of observations $\Omega_k \in \{0,1\}$ at fixed rate $\nu_\Omega= \Delta t^{-1},$ updating its posterior $b(\bm{x})$ after each incoming observation using Bayes' rule (Eq.~\ref{eq:bayes}). We assume that the agent performs the Bayesian update using the true \emph{single-time} likelihood $p_\Omega(\bm{x}) \equiv {\rm Pr}(\Omega |\bm{x}).$ That is, the agent does \emph{not} have knowledge of the underlying correlations, but otherwise has a perfect model.

To make progress, suppose the discrete observation sequence $\Omega_k$ satisfies the first-order Markov property
\begin{equation}
{\rm Pr} (\Omega_k| \Omega_{k-1}, \Omega_{k-2}, \dots, \Omega_{k-i}) = {\rm Pr} (\Omega_k| \Omega_{k-1}), 
\end{equation}
for any $i\in \mathbb{N}.$ This is a strong assumption which essentially says that the correlations decay exponentially in time. To be more precise, it is useful to introduce the  nonunit eigenvalue of the transition matrix for $\Omega_k$, which has entries $p_{\Omega' \Omega}.$ We will call this eigenvalue $\rho(\bm{x})$; an elementary calculation yields 
\be\label{eq:rho} \rho(\bm{x})= p_{11}(\bm{x})-p_{10}(\bm{x}) = \frac{C(\Delta t;\bm{x})}{C(0;\bm{x})}.\ee 
Here, we have introduced the two-time likelihood
\[
p_{\Omega'\Omega}(\bm{x}) = \mathrm{Pr}(\Omega(\bm{x},t+\Delta t) = \Omega' | \Omega(\bm{x},t )= \Omega )
\]
and the two-time function for the signal $\Omega(\bm{x},t)$
\begin{equation}
C(\tau;\bm{x}) = \langle \Omega(\bm{x},t+\tau) \Omega(\bm{x},t)\rangle_t - \langle \Omega(\bm{x},t)  \rangle_t^2.
\end{equation}

The first-order Markov assumption then predicts that the two-time function obeys
\[
C(t_k; \bm{x}) = C(0;\bm{x}) \rho^{k},
\]
for each $t_k = k \Delta t.$ This implies a correlation time $\tau_c=-\Delta t/\log |\rho|.$ However, note that this model is poor for describing anticorrelations, since it cannot capture the correct oscillation frequency (which it predicts to be either $\nu_\Omega$ or 0.)
In the stationary agent setting, the eigenvalue $\rho$ turns out to be equivalent to the Pearson coefficient between consecutive observations, which can be seen by applying the law of total probability $p_1 = p_{11} p_1 + p_{10} p_0$ to Eq.~\ref{eq:rho}. We therefore have $|\rho\le 1,$ with equality signaling perfect correlation (or anticorrelation for negative sign), while $\rho=0$ is equivalent to an absence of correlations \footnote{We will use $\rho$ throughout this work as a useful measure of correlation strength; note that it may be generalized to the moving agent setting by replacing $p_{11}$ and $p_{10}$ with their action-dependent analogs.}.

\begin{figure*}
    \centering
    \includegraphics[width=\linewidth]{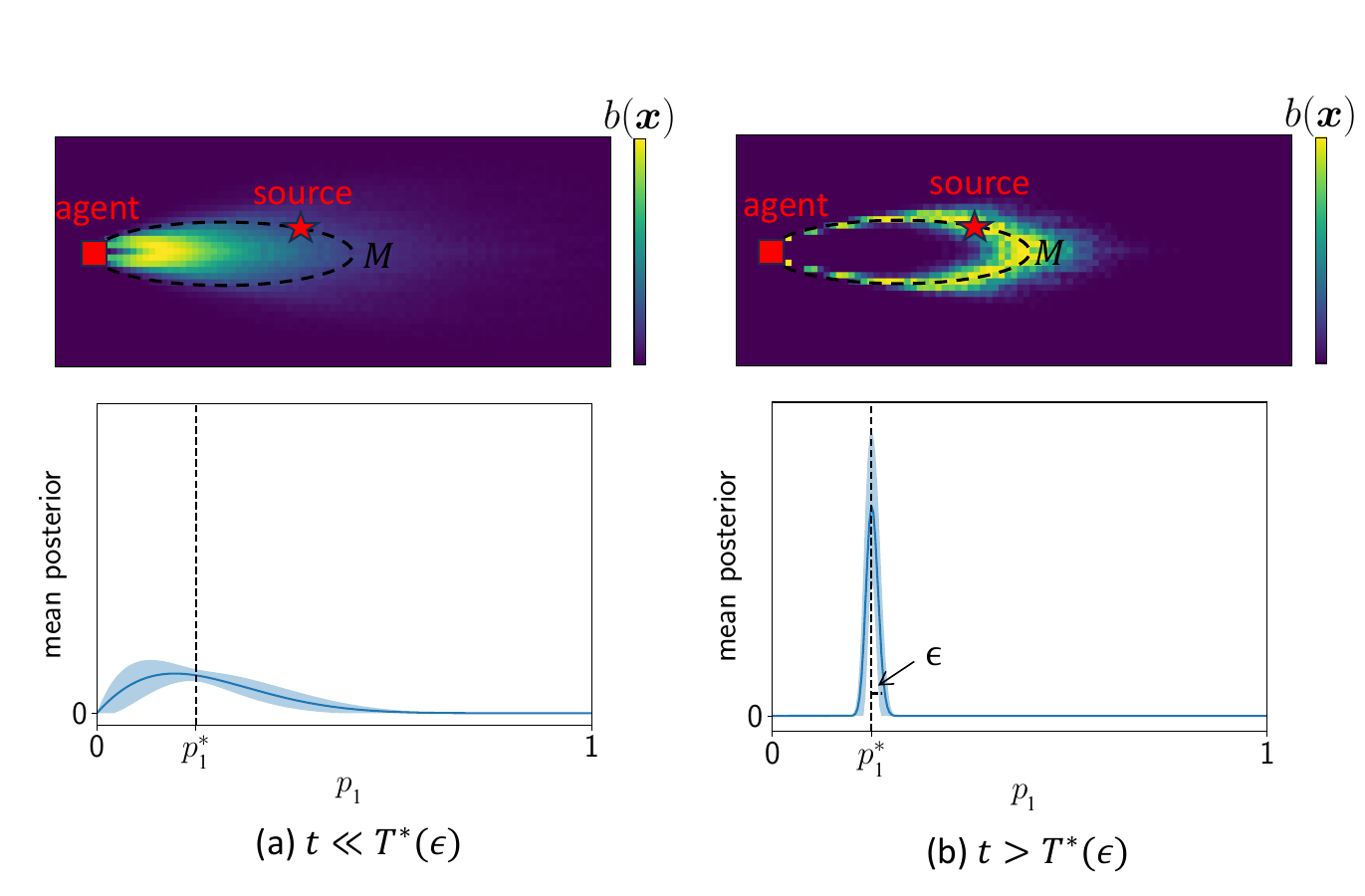}
    \caption{Cartoon illustrating the evolution of a stationary agent's posterior. The upper plots depict a typical posterior estimate of the source location $b(\bm{x})$. The lower plots show the ensemble average posterior estimate of the likelihood $p_1,$ with the shaded error bars indicating the standard deviation of this estimate over an ensemble of agents. [To be clear, this cartoon is intended to be conceptual and is not depicted to correct scale.] (a) At time $0<t \ll T^*$ only a few observations have been made, and the posterior on $p_1$ is noisy (not converged in the sense of probability) and is not centered on the ground truth $p_1^*$. There may still be significant memory of the prior (which was induced by an encounter at time zero). This corresponds to a posterior over physical space which is not peaked near the true source location. (b) After a time $t>T^*,$ the agent can estimate $p_1$ within a tolerance $\epsilon$ of $p_1^*$; outside this tolerance, the posterior is converged in probability to zero. This corresponds in space to a posterior peaked at the level curve of equal likelihood $M=\{\bm{x} : p_1(\bm{x}) = p_1^*\},$ which contains the true source location. In the na\"ive strategy, the agent only begins to move when time $T^*$ has elapsed, and thereafter explores the support of $b.$ In Appendix \ref{app:theory}, we derive an expression for the appropriate $T^*.$}
    \label{fig:stationary_agent}
\end{figure*}

In spite of the limitations of the first-order Markov model, under its assumptions we can establish an upper bound on the typical time $T^*$ for the stationary Bayesian agent to estimate the source position (see Appendix \ref{app:theory}). The result is 
\be \label{eq:tstar2}
T^* \le A \nu_\Omega^{-1} \max \left\{\frac12, \frac{1+\rho}{1-\rho}\right\},
\ee
where $A$ is a constant that depends on the desired precision of estimation. The expression for $T^*$ comes from taking the maximum of two distinct characteristic times: first, the time for the posterior to be sharply peaked at the ground truth, and second, the time for the posterior to be converged in probability (i.e., in an ensemble sense). Note that for a single stationary agent making binary observations, it is impossible to localize the source beyond a level set of codimension one---see Fig.~\ref{fig:stationary_agent}.

Two interesting observations are worth mentioning. For one, the upper bound diverges when $\rho \to 1$: if the observation sequence is correlated on very long time-scales, it is impossible to estimate the single-time likelihood $p_1$ in finite time due the presence of arbitrarily long chains of encounters. For another, if $\rho<0,$ the upper bound is less than if $\rho=0$; whereas positive correlations slow down convergence of the posterior, anti-correlations \emph{speed} up the convergence and \emph{help} the agent localize the source more quickly, even when the agent doesn't know the signal is anti-correlated. 

In a real flow, due to the tendency of passive scalars to organize into filaments and other spatial structures, we generally expect successive encounters with the scalar to be positively correlated over sufficiently small separation distances; a Markovian model therefore seems appropriate far downwind.
On the other hand, in the vicinity of the source, and in the presence of a sufficiently strong mean flow, passive scalars have not had time to separate and are bound tightly into a coherent beam. If an agent moving crosswind passes through the beam at speed $v_A$ and makes an encounter, subsequent encounters will occur with small probability since the agent will exit the beam. This is effect will tend to anti-correlate subsequent observations with the encounter. In fact, the coherent structure of this beam persists, albeit on larger scales, as it moves downwind, as can be visualized in the snapshots of Fig.~\ref{fig:search_cartoon}. We will see that large-scale coherence can have important and generally helpful effects on olfactory search.

\section{Methods}\label{sec:methods}
\subsection{The DNS}
We solved the incompressible 3D Navier-Stokes equations 
\begin{align}
\partial_t \bm{u} + (\bm{u} \cdot \nabla) \bm{u} &= - \nabla p + \nu \nabla^2 \bm{u} + f, \\
\nabla \cdot \bm{u} &= 0
\end{align}
using a pseudospectral method on a $1024 \times 512\times 512$ grid, with periodic boundary conditions. The simulation gridspacing was $\delta x= \delta y = \delta z \simeq \eta,$ where $\eta$ is the Kolmogorov length. Here $f$ is a random isotropic forcing at large scales, with correlation time $\simeq 1.2  \tau_\eta$ (where $\tau_\eta$ is the Kolmogorov time-scale). For all simulations, the Reynolds number at the Taylor scale was $\mathrm{Re}_\lambda \simeq 150,$ which is in the turbulent regime. Five point sources were moved at constant velocity $U \hat{x}$ for $U/u_{\rm rms}\in\{ 0,1.2,2.5,4.9,7.4\}.$ The sources each emitted 1000 particles in a small radius every 10 simulation timesteps, or about every $1/15 \tau_\eta.$ The particles were advected by the flow according to the tracer equation
\[  \dot{\bm{X}} = \bm{u}(\bm{X},t)\]
and their positions $\bm{X}$ were tracked and dumped at every $\tau_\eta.$ Finally, the Galilean transformation $\bm{x} \to \bm{x} - Ut \hat{x}$ was applied to the data, thus returning to the frame where the sources are stationary and there is a mean flow $-U \hat{x}.$ All statistical data were averaged over the five sources, which were treated as independent.

\subsection{POMDP details}\label{sec:pomdp}
Generically, POMDP formalism \cite{astrom1965,kaelbling1998} describes an agent interacting with a stochastic, Markovian environment by taking actions. The system evolves according to some ${\rm Pr} (s'|s,a),$ where $s,s'$ are system states and $a$ is the agent action. The agent maintains a posterior $b(\bm{s})$ over the possible system states and makes observations $\Omega$ which are generally governed by some likelihood ${\Pr}(\Omega|s,a)$. The observations are used to update the posterior, and the agent seeks to take actions in such a way that minimizes over time some cost function $\mathcal{C}(s,a),$ where $s$ is the current state and $a$ the action just taken. In our problem, the observations are encounters with the passive scalar cue, the states are the position of the agent relative to the source, the actions are motion in a cardinal direction, and the cost should be chosen to penalize long arrival times (or equivalently, reward short arrival times). 

\begin{figure*}
    \centering
    \includegraphics[width=0.9\linewidth]{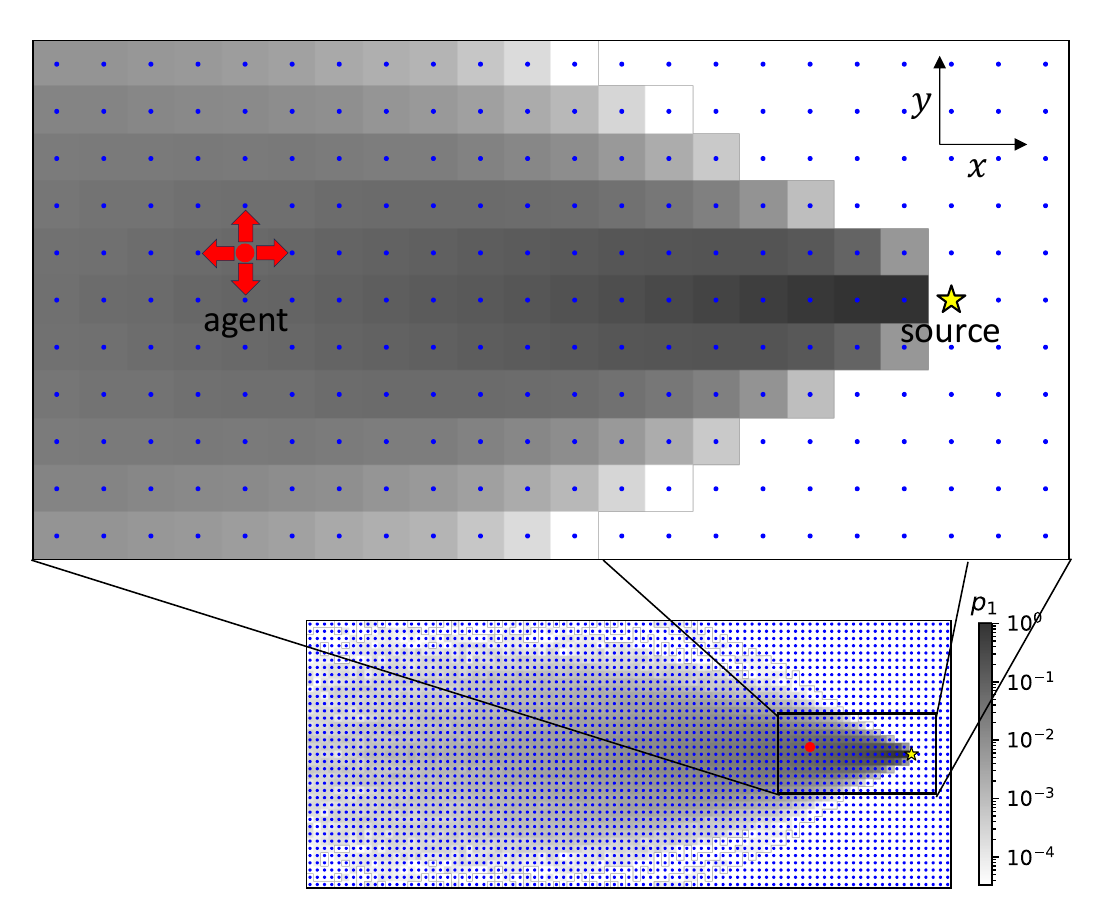}
    \caption{Schematic illustrating the POMDP for a particular choice of the mean wind. The agent is shown in red, and the source as a yellow star. The states of the system (blue points) are, collectively, a chosen grid of possible positions of the agent with respect to the source. We denote the current state by $\bm{x}$. At each time step, the agent makes an observation $\Omega \in\{0,1\},$ where $0$ represents no encounter and 1 an encounter. The likelihood of an encounter $p_1(\bm{x})$ is shown in grayscale. After making an observation, the agent takes an action $\bm{a}$ by moving to one of the four adjacent grid points; the four possible actions  are shown as purple arrows. The prior probability at the initial time of an agent at a given position is proportional to $p_1.$}
    \label{fig:pomdp-schematic}
\end{figure*}

In our setup (see Fig.~\ref{fig:pomdp-schematic}), the search is performed on a quasi-2--D slab at constant $z$ which contains the source. The slab consists of an array of cubical cells with side length $\simeq 15 \eta$. The total number of cells in the slab was held approximately fixed at $\approx 3250$, but the aspect ratio was varied depending on the wind speed to accommodate different plume shapes. The grid sizes are shown in Fig.~\ref{fig:likelihoods}.

In order to translate this problem into a POMDP, at each $\Delta t = \tau_\eta,$ the number of particles in each cell was counted in order to obtain a concentration field $c(\bm{x})$. The agent begins at some chosen point and at each time step makes an observation $\Omega$---either an encounter or no encounter---then moves to one of the four adjacent grid cells. The actions can thus be considered vectors $\bm{a}\in\{(0,1),(1,0),(-1,0),(0,-1)\}.$ The agent is assumed to be a strong enough swimmer or flier that the motion is deterministic and the agent is not advected by the flow, i.e., ${\rm Pr}(s'|s,a) = \delta_{s',s+a}$. Unless otherwise stated, we set the threshold $c_{\rm thr}$ to $200$, in units of the number of particles.

The posterior of the agent $b(\bm{x})$ is a probability distribution on the grid representing the possible positions of the agent, relative to the source. At each time step, $t_k$, after the agent makes observation $\Omega$ and takes action $\bm{a}$, the posterior is updated according to Bayes' theorem (\ref{eq:bayes}) adapted to our set-up:
\be
b_k(\bm{x}| \Omega,\bm{a} ) = \frac{p_\Omega(\bm{x}-\bm{a}) b_{k-1}(\bm{x}-\bm{a}) }{\sum_{\bm{x}'} p_\Omega(\bm{x}') b_{k-1}(\bm{x}') }. 
\ee
 As more observations are made, the posterior, which contains all the agent's current information about the source location, becomes more informative. An important point is that the prior is chosen to be the normalized likelihood of an encounter, $b_0(\bm{x}) \propto p_1 (\bm{x}).$ This models the assumption that the agent only begins searching after encountering the passive scalar. For self-consistency, the true position of the source (relative to the agent) is drawn at time $t_0=0$ from the prior, as in Refs.~\cite{loisy2022,loisy2023,panizon2023,heinonen2024}; the agent always starts in the same position in the gridworld (see caption of Fig.~\ref{fig:likelihoods}). When searching in the DNS, this is realized in practice by choosing a random encounter event from the entire time series data.
 
\subsection{Correlation-aware agent}\label{sec:aware}
The POMDP previously described involves an agent that is unaware of correlations, and updates its posterior with the single-time likelihood. Suppose instead that the agent is aware that the system is correlated and has access to the action-dependent two-time likelihoods 
\[p^{\bm{a}}_{\Omega' \Omega}(\bm{x}) \equiv {\rm Pr}\big(\Omega(\bm{x},t+\Delta t)=\Omega'\big|\Omega(\bm{x}-\bm{a},t) = \Omega\big).\]
The agent can then exploit this information by updating its posterior with the two-time likelihood instead of the single-time likelihood. 

The resulting decision process can still be formalized as a POMDP by augmenting the state space to include the observation. The state transition probability is now set by the two-time likelihood, i.e.,
\[ {\rm Pr}(\bm{x}',\Omega'|\bm{x},\bm{a}, \Omega) = p^{\bm{a}}_{\Omega'\Omega}(\bm{x}') \delta_{\bm{x}',\bm{x} + \bm{a}},\]
where observation $\Omega'$ is made at position $\bm{x}'$ and observation $\Omega$ is made at $\bm{x}.$ Meanwhile, the belief is still just a distribution over $\bm{x}$ since $\Omega$ is fully observable, and the Bayesian update rule is now 
\[
b_k(\bm{x}|\Omega',\Omega,\bm{a}) = \frac{ p_{\Omega' \Omega}^{\bm{a}} (\bm{x}) b_{k-1}(\bm{x}-\bm{a})}{\sum_{\bm x'} p_{\Omega' \Omega}^{\bm{a}} (\bm{x}') b_{k-1}(\bm{x}'-\bm{a}) }.\]
 We will solve this POMDP in order to obtain first-order ``correlation-aware'' policies. In principle, this kind of approach could be extended to higher-order Markov processes, but at exponentially growing computational cost, and with increasingly poor empirical estimates of the multi-time likelihoods.

\subsection{Optimal policy}
The choice of action for a given posterior is expressed by the policy $\pi: b\mapsto \bm{a}$. The POMDP admits a unique optimal policy which minimizes, in expectation, the arrival time to the source $T.$ We define the cost function to be a unit penalty at each time step until the agent finds the source, i.e., \ 
\be
\mathcal{C}(\bm{x},\bm{a}) = \mathcal{C}(\bm{x}) =
\begin{cases}
0, &\bm{x}= 0 \\
-1,& \bm{x}\ne 0,
\end{cases}
\ee
so that the total cost is the arrival time $T$ (recall that $\bm{x}$ is the agent's position relative to the source, so $\bm{x}=0$ means the agent has arrived). In practice, we will introduce a discount factor $\gamma\in(0,1)$ which regularizes the problem by reducing the influence of long time horizons.  We then seek a policy which minimizes the expectation of the total (discounted) cost $\mathcal{C}_1 + \gamma \mathcal{C}_2 + \gamma^2 \mathcal{C}_3+\dots= \frac{1-\gamma^T}{1-\gamma}$, where $\mathcal{C}_t$ is the reward received at timestep $t,$ conditioned on the current posterior and the policy; this quantity is called the value function $V^\pi[b]$. The value corresponding to the optimal policy, $V^*,$ is the solution to the so-called the Bellman optimality equation or dynamic programming equation \cite{sondik1978}. In the case where the agent is unaware of correlations, the Bellman equation for this problem reads
\begin{widetext}
\be\label{eq:bellman}
V^*[b] =\begin{cases} 0,& b=\delta(\bm{x}) \\ 1 + \gamma \min_{\bm{a}} \sum_{\Omega,\bm{x}} p_\Omega(\bm{x}+\bm{a})b(\bm{x}) V^*[b(\cdot|\Omega,\bm{a})], & \textrm{otherwise.}\end{cases} 
\ee
\end{widetext}
Note that in this model, the agent is able to observe when it arrives to the source, collapsing its belief to a delta function; when this is the case, the value is zero since the time to the source vanishes. A similar Bellman equation can be written down in the correlation-aware case:
\begin{widetext}
\be\label{eq:bellman2}
V^*[b,\Omega] =\begin{cases} 0,& b=\delta(\bm{x}) \\ 1 + \gamma \min_{\bm{a}} \sum_{\Omega',\bm{x}} p_{\Omega'\Omega}^{\bm{a}}(\bm{x}+\bm{a})b(\bm{x}) V^*[b(\cdot|\Omega,\bm{a}),\Omega'], & \textrm{otherwise,}\end{cases} 
\ee
\end{widetext}
where now $V^*$ depends on both the belief and the last observation made.

We approximately solve for $V^*$ by applying the SARSOP \cite{sarsop} algorithm to the empirical likelihood data, using $\gamma=0.98,$ as close as possible to unity without compromising the convergence of the algorithm. The quasi-optimal policy can then be identified, for each $b$ (and $\Omega,$ in the correlation-aware case) as the action that realizes the min in the RHS of Eq.~\ref{eq:bellman}.  For further technical details on the POMDP implementation and solution, we refer the reader to Refs.~\cite{loisy2022,heinonen2023,loisy2023}.

The SARSOP policy continues to refine the policy indefinitely, until terminated by the user. We chose to compute all policies (both correlation-unaware and -aware) over a fixed walltime of 4000 seconds. The performance of the policy tends to fluctuate somewhat with continued iteration, and therefore, when we plot arrival time statistics in Sec.~\ref{sec:results}, there is some uncertainty in the results associated with uncertainty in the policy itself; this will not be reflected explicitly in the error bars.

\subsection{Infotaxis and SAI}
As an inexpensive alternative to optimization, one can choose a heuristic strategy, the best-known of which is infotaxis (and its variants). Infotaxis \cite{infotaxis} seeks to minimize the Shannon entropy of the posterior $b$; at each time step the agent selects the action
\[ \pi_{\rm info}[b] = \argmin_{\bm{a}} \sum_{\Omega,\bm{x}} p_\Omega(\bm{x}+\bm{a})b(\bm{x}) H[b(\cdot|\Omega,\bm{a}))],  \]
where $H[b] = - \sum_s b(s) \log b(s).$ [We break ties between actions by random choice.]

Infotaxis prioritizes exploration (gathering information about the source) at the cost of exploitation (attempting to strike to the source based on the information already gathered), and for this reason is clearly suboptimal in some cases. One way to improve infotaxis is to try to simultaneously minimize both the entropy \emph{and} the expected distance to the source. This leads to space-aware infotaxis (SAI) \cite{loisy2023}. While SAI was originally envisioned as a single policy, we will presently view SAI as a one-parameter family of policies
\begin{widetext}\be \label{eq:sai1} \pi_{{\rm SAI},\lambda}[b] = \argmin_{\bm{a}} \sum_{\Omega,\bm{x}} p_\Omega(\bm{x}+\bm{a})b(\bm{x}) J_\lambda[b(\cdot|\Omega,\bm{a}))],
\ee\end{widetext}
where 
\be \label{eq:sai2} J_\lambda[b] = \log \left( (1- \lambda) e^{H[b]} + \lambda \sum_{\bm{x}} \norm{\bm{x}}_1 b(\bm{x})\right),  \ee
$\bm{x}$ is the agent position, $\bm{y}$ is the source position, $\norm{\cdot}_1$ is the $L_1$ norm, and $\lambda$ is a parameter which controls the degree of exploitation vs.\ exploitation. The objective function of the policy, $J,$ nonlinearly interpolates between the entropy of the belief and the expected distance to the source. Therefore, $\lambda=0$ recovers infotaxis (pure exploration), while $\lambda =1 $ yields a greedy (exploitative) policy which tries to strike toward the source; any $\lambda \in (0,1)$ interpolates between the two.

Viewing infotaxis as a special case of this family, we will mainly show results for the empirically best SAI policy (i.e., choice of $\lambda$) for a given task.

\section{Results}\label{sec:results}
\subsection{Statistics of encounters with the passive scalar}
\begin{figure*}
    \centering
    \includegraphics[width=\linewidth]{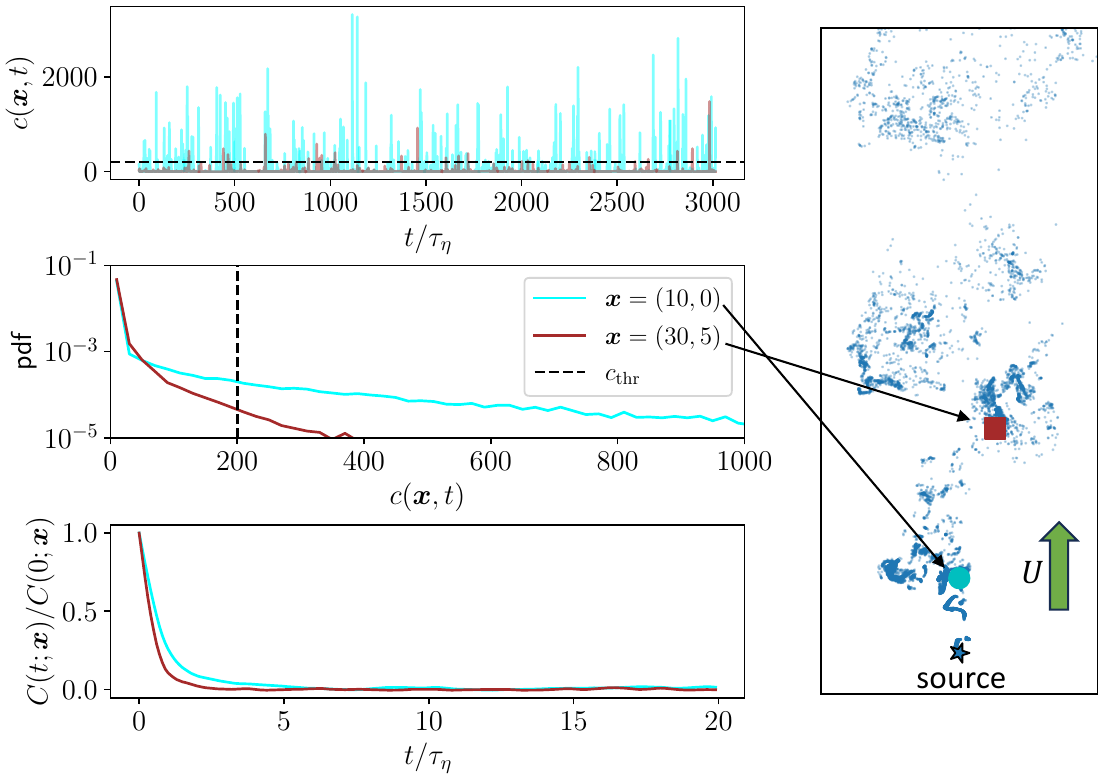}
    \caption{Concentration time series, concentration pdf, and two-time function $C(\bm{x},t)$ for the observation signal $\Omega(\bm{x},t)=\theta(c(\bm{x},t)-c_{\rm thr})$ with $c_{\rm thr} = 200,$ evaluated at two test points, for the case $U=2.5 u_{\rm rms}.$ The positions of the test points are indicated at right within a snapshot of the DNS which shows Lagrangian particle positions in blue.}
    \label{fig:timeseries}
\end{figure*}

First, in Fig.~\ref{fig:timeseries} we visualize the concentration time series at two test points for the case $U=2.5 u_{\rm rms}$, and compare it to our choice of threshold $c_{\rm thr}=200.$ In the same figure, we evaluate the pdf of the concentration field as well as the two-time function $C(t;\bm{x})$ for the observation sequence. This figure highlights that encounters, as defined, are rare events and that the statistics are nonstationary as the agent moves through the flow.

\begin{figure}
    \centering
    \includegraphics[width=\linewidth]{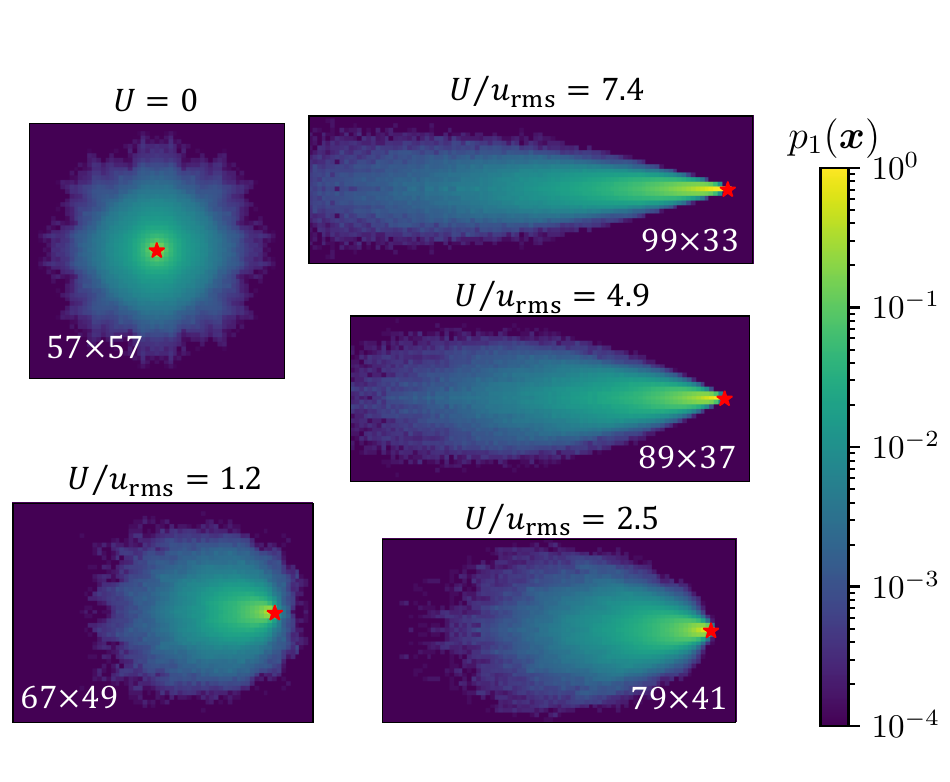}
    \caption{Empirical observation likelihoods for $c_{\rm thr} = 200,$ counterclockwise from top left in order of increasing wind speed. The sizes of the gridworlds are shown with the correct relative scale, and the gridworld dimensions are indicated in the corner of each plot. The source is shown as a red star in each gridworld. In order of increasing wind speed, the agent started the search at positions $(28,28),$ $(10,24),$ $(4,20),$ $(4,18),$ and $(4,16).$}
    \label{fig:likelihoods}
\end{figure}

Next, we show the empirical observation likelihoods obtained from the DNS data, for threshold $c_{\rm thr}=200$ (Figs.~\ref{fig:likelihoods}). These are obtained by time averaging, averaging independently over the five sources, and symmetrizing \footnote{In the windy case, we invoke $\mathbb{Z}_2$ symmetry with respect to y-reflections, and in the isotropic case we symmetrize over the largest subgroup of the plane rotations $O(2)$ which respects the grid, i.e., the square dihedral group $D_4$}. For strong mean flow, the likelihoods show strong conical structure, vanishing upwind of the source and decaying rapidly as one moves away from the centerline, in agreement with predictions from Ref.~\cite{celani2014}. As $U$ decays the cone where encounters are probable becomes more ovate and eventually gains some support upwind of the source. In the isotropic case, the empirical likelihoods are approximately rotationally invariant and decay as a function of distance to the source. 


\begin{figure*}
    \centering
    \includegraphics[width=\linewidth]{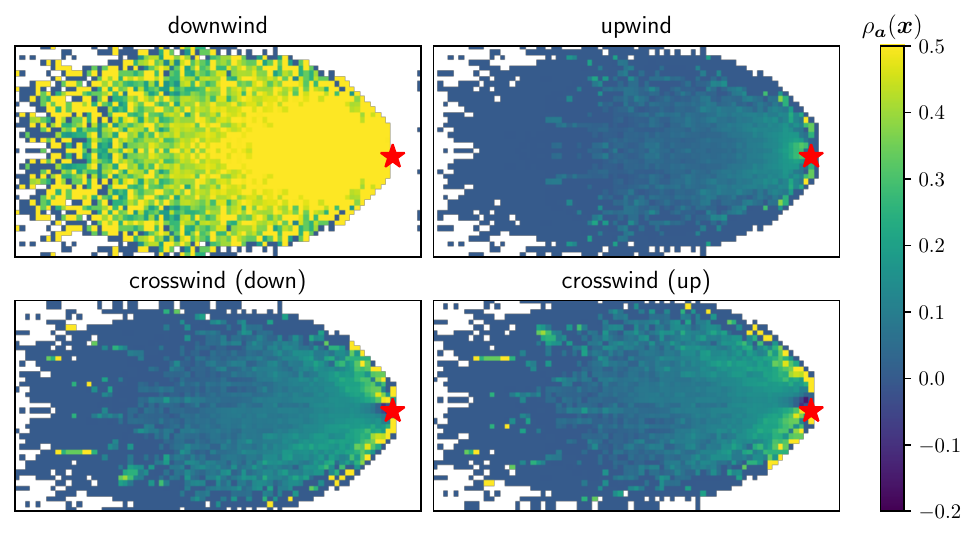}
    \caption{Heatmaps showing the one-step correlation strength $\rho_{\bm a}$ as a function of position for each of the four possible actions, when $c_{\rm thr}=200$ and $U/u_{\rm rms}=2.5.$ The correlations are strong when the agent moves downwind (top left panel), which negates the decorrelating effect of the sweeping by the mean flow.}
    \label{fig:v3_deltas}
\end{figure*}

These likelihoods contain no information about correlations. As an illustrative example, in Fig.~\ref{fig:v3_deltas} we show, for $c_{\rm thr} =200$ and $U\simeq 2.5 u_{\rm rms},$ the action-dependent correlation strength \[\rho_{\bm{a}}(\bm{x}) \equiv p_{11}^{\bm{a}}(\bm{x}) - p_{10}^{\bm{a}}(\bm{x}),\]
where $p_{\Omega' \Omega}^{\bm{a}}(\bm{x}) = \mathrm{Pr}(\Omega(\bm{x}+\bm{a},t+\Delta t)=\Omega'| \Omega(\bm{x},t)=\Omega)$ is the conditional probability when taking action $\bm{a}.$ The correlations are mostly positive and generally decay with distance from the source. One also immediately sees how, in the presence of a mean flow, the downwind action exhibits the strongest correlations; this is because that action counteracts the decorrelating effect of sweeping by the mean flow.

\begin{figure*}
    \centering
    \includegraphics[width=\linewidth]{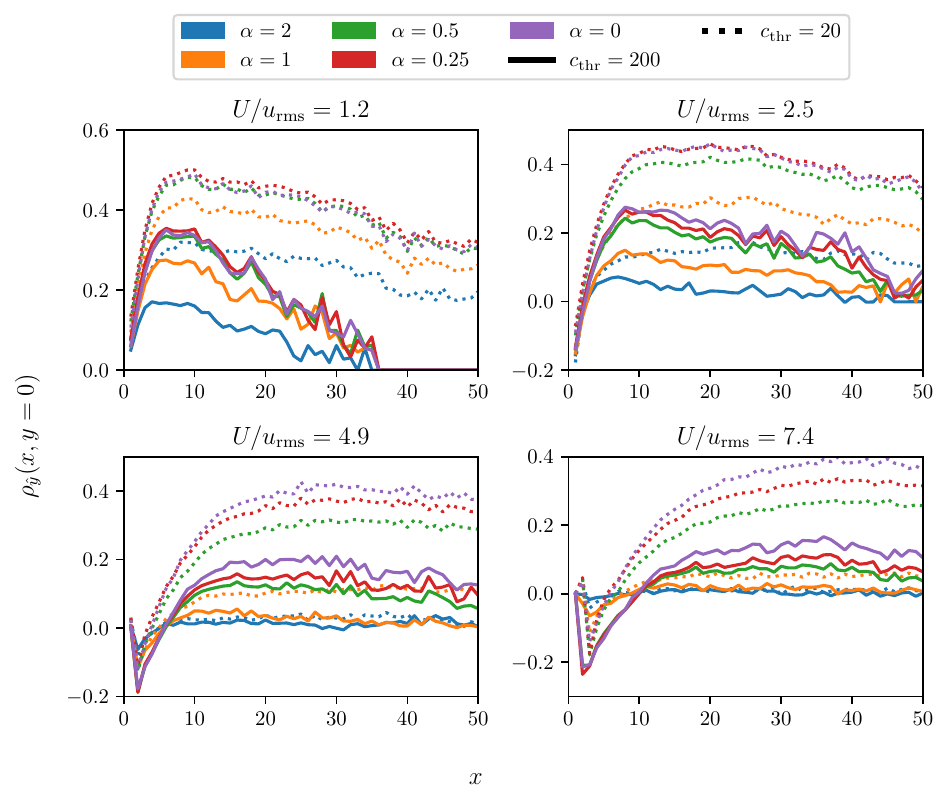}
    \caption{Correlation strength $\rho$ for the action of moving crosswind from the centerline in the windy setting, shown as a function of the downwind distance $x$ for several choices of threshold and the time rescaling parameter $\alpha \equiv \tau_\eta/\Delta t.$ Note that correlations tend to get stronger as $\alpha \to 0$ or $c_{\rm thr} \to 0$ and slowly decay as $x\to \infty.$ Also note the presence of anticorrelations close to the source when $U\ge 2.5 u_{\rm rms}.$}
    \label{fig:windy-deltas}
\end{figure*}

Correlation strengths depend sensitively on many parameters: spatial position, the action taken, the threshold, the mean flow, and the rate of observation.  To partially illustrate these dependencies, we plot in Fig.~\ref{fig:windy-deltas} the value of $\rho$ for a particular action (moving crosswind from the centerline), when $U\ne 0$. The rate of observation is altered by a rescaling parameter $\alpha \equiv \tau_\eta/\Delta t$, where $\Delta t$ is the timestep at which the agent moves and makes observations; this transformation will later be used to tune the correlations without altering the single-time likelihood. We see that, as claimed in the introduction, choosing a higher threshold tends to suppress correlations, and slowing time ($\alpha \to 0$) tends to intensify them. To be clear, $\alpha=1$ means the timestep between snapshots is $\tau_\eta.$ The correlations also decay with downwind distance (or distance from the source in the isotropic case), albeit slowly. As previously noted, encounters are \emph{anti-correlated} when moving crosswind close to the source, an effect which is systematically underestimated when measuring $\rho_{\bm a}$ alone.

\subsection{Arrival time statistics}\label{sec:stats}
Now we study the statistics of arrival times to the source under various flow conditions and policies by performing search trials in the DNS data. As mentioned previously, for each trial, we draw the initial position of the agent with respect to the source by selecting at random one position and timestep from the time series data where the local concentration was above threshold. This is consistent with the agent's prior, which is proportional to the single-time likelihood. We will present results for $c_{\rm thr} =200.$

The search trials are conducted for at most $T_{\max}=2500$ timesteps. For a given trial, there is a small but finite probability that the agent will not find the source within this time, which we refer to as ``failure.'' This means that average arrival times will be infinite unless we condition on not having failed, but this in turn can be misleading if the failure rate is relatively large. Instead, we choose to renormalize the arrival times in the following way:
\be \label{eq:renorm}
T \to F(T) \equiv \sum_{t=1}^{T} \beta^{t-1} = \frac{1-\beta^{T}}{1-\beta},
\ee
with $0<\beta<1$. This cuts off large arrival times past a horizon $H = 1/\log \beta^{-1} \simeq 1/(1-\beta),$ so that if $T \gg H,$ then $F(T) \simeq H.$  As $\beta \to 1$ we recover the unrenormalized arrival time $T$. If $T_{\max}$ was saturated, then the arrival time was treated as infinite. We chose $\beta=0.999,$ corresponding to $H=1000,$ which is typically deep into the tail of the arrival time pdf. Physically, the horizon might be thought of as a time beyond which arrival to the source provides no benefit. In what follows, all mean arrival times have been computed after renormalization.

We will usually show the excess arrival time $T- T_{\rm MDP},$ where $T$ is the total arrival time and $T_{\rm MDP}$ is the minimum possible time, i.e.,\ the distance to the source---so named because it is the optimum of the underlying fully observable Markov decision process. The excess arrival times are then normalized by $T_{\rm MDP}.$

\subsubsection{Experiments varying the measurement frequency: the unaware agent} \label{sec:unaware_exp}
We first show the results for an agent trained ignoring correlations, i.e., solving the Bellman equation (\ref{eq:bellman}) and comparing its performances against the infotactic heuristics. To adjust the strength of correlations (while fixing the single-time likelihood of an encounter), we use the previously defined scaling parameter $\alpha \equiv \tau_\eta/\Delta t$, where $\Delta t$ is again the timestep at which the agent moves and makes observations, while fixing $\Delta x = v_A \Delta t$. Hence, under $\alpha \to \alpha'$ we send $v_A \to v_A \alpha /\alpha'$ and $\nu_\Omega \to \nu_\Omega \alpha/\alpha'.$ We chose several values of $\alpha\in\{0,1/4,1/2,1,2,\infty\}$ by which to rescale $\Delta t.$ To be clear, $\alpha=0$ means the agent searches in a frozen concentration field, and $\alpha=\infty$ means the flow is completely uncorrelated. In order to obtain the $\alpha=1/4,1/2$ points, the raw Lagrangian data were interpolated at intermediate times using the particles' instantaneous velocities and accelerations; this was then coarse-grained into a concentration field with finer resolution in time. We roughly expect the correlations to increase in strength as $\alpha \to 0,$ since this eliminates the decorrelating effect of sweeping by the mean flow. On the other hand, for $\alpha=\infty,$ observations were drawn randomly from the likelihood rather from DNS data. 

\begin{figure*}
    \centering
    \includegraphics[width=\linewidth]{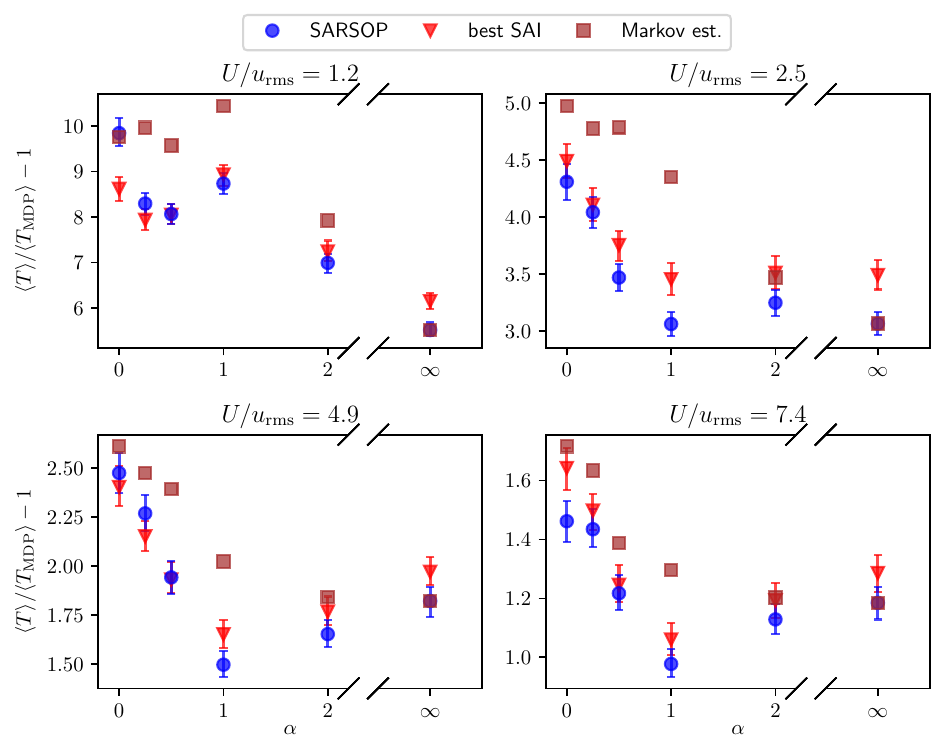}
    \caption{Mean excess arrival times to the source, normalized by minimum mean time, as function of the time rescaling parameter $\alpha,$ for $U\ne0.$ Error bars indicate 95\% confidence levels on the means and were obtained by bootstrapping. The brown squares indicate the first-order Markov prediction of Eq.~\ref{eq:markov_est}.}
    \label{fig:windy_mreat_100}
\end{figure*}

\begin{figure}
    \centering
    \includegraphics[width=\linewidth]{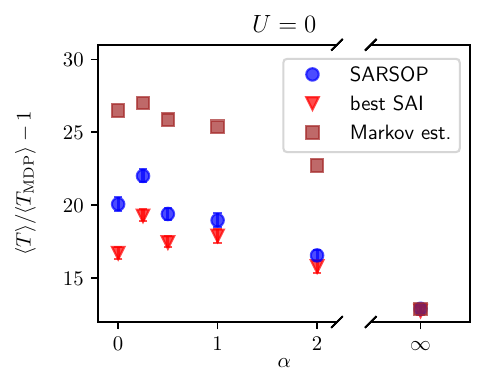}
    \caption{Mean excess arrival times to the source as function of the time rescaling parameter $\alpha,$ for the isotropic case $U=0.$}
    \label{fig:isotropic_mreat_100}
\end{figure}

The mean excess arrival times, normalized by mean minimum arrival time, are plotted as a function of $\alpha$ in Figs.~\ref{fig:windy_mreat_100}--\ref{fig:isotropic_mreat_100}, for both with and without mean wind. For each flow speed and each $\alpha$, we show the results for both the optimized SARSOP policy and the best SAI policy. We tested each $\lambda=j/10$  for $1\le j \le 10,$ where $\lambda$ is the SAI parameter which sets the exploitation/exploration tradeoff---see Eqs.~\ref{eq:sai1}--\ref{eq:sai2}. 

Na\"ively, one may expect the arrival time to decay monotonically as $\alpha$ increases. However, the action dependence of $\rho$ can have complicated effects, notably for $U=1.2 u_{\rm rms}$ (top left panel of Fig.~\ref{fig:windy_mreat_100}). The correlation strength, $\rho$, is generally largest for the downwind action, as this motion counteracts the decorrelating effect of sweeping by the mean flow (see, for example, Fig.~\ref{fig:v3_deltas}). On the other hand, the sweeping effect is mitigated as $\alpha \to 0,$ which complicates the $\alpha$-dependence of $\rho$ for the downwind action: in particular, $\bar{v}$ has a local maximum when $\alpha=2u_{\rm rms}/U,$ vanishes at $\alpha=u_{\rm rms}/U,$ and diverges at $\alpha=0$ (see Eq.~\ref{eq:dimensionless}), resulting in strongly non-monotonic dependence of $\rho$ on $\alpha.$ As the mean wind decreases, the agents opt to move downwind more frequently, so that for $U=1.2 u_{\rm rms}$ the downwind action strongly influences $T^*$, resulting in non-monotonic dependence of $T^*$ on $\alpha.$  

Another notable feature of Fig.~\ref{fig:windy_mreat_100} is that for large mean flow and $\alpha \sim 1$, the mean arrival time is actually \emph{smaller} than in the uncorrelated case,  $\alpha = \infty$. We have checked that there is roughly uniform improvement in arrival time relative to the uncorrelated policy, regardless of starting distance to the source. We have also checked that if the observations are drawn stochastically from the one-step conditional likelihoods $p_{\Omega' \Omega}^{\bm{a}}(\bm{x})$ instead of from the DNS--- hus suppressing all but short-range correlations---then this effect disappears. Therefore, it appears that \emph{long-range correlations} can improve the agent's arrival time to the source. We will examine this more closely in Sec.~\ref{sec:discussion}.

From the general trends in Fig. \ref{fig:windy_mreat_100}, we can conclude that in the presence of mean wind and small correlations, the SARSOP policy for agents unaware of correlations generally outperforms the SAI heuristic, according to its quasi-optimality with respect to the uncorrelated problem. Exceptions occur when the correlations are strong due to $\alpha$ being small. However, we emphasize that getting the most out of SAI requires carefully optimizing the parameter $\lambda.$

In Fig. \ref{fig:windy_mreat_100} we also compare these results to a semi-empirical estimate derived from the upper bound  (\ref{eq:tstar2}) obtained for the case when the agent is fixed in space while updating its belief, discussed in  Sec.~\ref{sec:theory}.

In order to apply Eq.~\ref{eq:tstar2} to our case, we need to take into account the fact that the agent is not static and that the correlations depend on the direction of motion. To do that, we suppose that the primary effect of the motion is to perform a spatial average on the trajectory, and we estimate the excess arrival time as an average of the static upper bound Eq.~\ref{eq:tstar2} over the joint probability $\mathrm{Pr}(\bm{x},\bm{a})$ of taking action $\bm{a}$ at position $\bm{x} $, and using the empirically value of the action-position dependent correlation parameters, $\rho_{\bm{a}}(\bm{x})$:
\begin{widetext}\be
\label{eq:markov_est}
\langle T  - T_{\rm MDP} \rangle \simeq \tilde T_{\rm uncorr} \sum_{\bm{a}, \bm{x}} \mathrm{Pr}(\bm{x},\bm{a}) \max\left\{\frac12,\frac{1+\rho_{\bm{a}}(\bm{x})}{1-\rho_{\bm{a}}(\bm{x})}\right\}.
\ee
\end{widetext}
where the overall normalization factor  $\tilde T_{\rm uncorr} = \langle T(\alpha=\infty) - T_{\rm MDP} \rangle $ is the excess arrival time in the absence of correlations
\footnote{These were obtained by performing search trials in uncorrelated data, while recording the actions taken at each position. Assuming that the action probabilities do not change much due to correlations, this allows us to compute the joint probability ${\rm Pr} (\bm{x},\bm{a}),$ and $\tilde T_{\rm uncorr}$ is obtained as the mean excess arrival time.},
and where ${\rm Pr} (\bm{x},\bm{a})$ and $\rho_{\bm{a}}(\bm{x})$ were both smoothed in space using a window averaging filter. As one can see, the first-order Markov estimate (Eq.~\ref{eq:markov_est}) captures the non-monotonicity shown for the case $U=1.2 u_{\rm rms}$ reasonably well, and agrees qualitatively with the observed arrival time performance. However, it tends to be pessimistic and systematically overestimates the arrival time in almost every case. Similarly, the estimate (\ref{eq:markov_est}) misses the minimum at $\alpha \sim 1$ shown by all the other windy cases, consistent with the claim that this effect was due to long-range correlations.

Concerning the isotropic case $U=0$ (Fig.~\ref{fig:isotropic_mreat_100}), where turbulent correlations are the largest because of the absence of sweeping, SAI outperforms SARSOP for all five finite values of $\alpha$ we tested; further, pure infotaxis ($\lambda=0$) is the best SAI policy in three out of five cases. One can interpret this result as follows: the strong correlations in the isotropic problem mean that the uncorrelated likelihood is far from the true likelihood which has been conditioned on the time history. This means a greedy policy, reflecting confidence in the information stored in the posterior, should be avoided, and thus optimizing the POMDP or taking the SAI greediness parameter $\lambda$ large are poor choices. In particular, we are led to a very different conclusion than that of \cite{loisy2022}, which found that SAI was close to optimal on the isotropic problem, outperforming pure infotaxis in particular---in fact, this is not generally true except in the absence of correlations. This result is similar in spirit to previous research \cite{heinonen2023} finding infotaxis to be the top performing policy when the model likelihood is strongly misspecified.

\subsubsection{Experiments varying the measurement frequency: the aware agent}\label{sec:aware_exp}
We now repeat the experiment of the previous subsection, but using an aware agent, obtained by using SARSOP to solve the POMDP described in Sec.~\ref{sec:aware} and Eq.~\ref{eq:bellman2}.

\begin{figure}
    \centering
    \includegraphics[width=\linewidth]{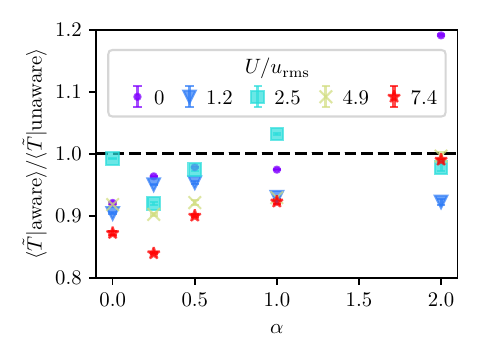}
    \caption{Mean excess arrival time as function of $\alpha$ for all choices of $U,$ using first-order correlation-aware SARSOP policies, plotted as the ratio to the corresponding result using an unaware policy. The dashed horizontal line indicates unity, i.e., no change in performance.}
    \label{fig:mreat_with_corr}
\end{figure}

In Fig.~\ref{fig:mreat_with_corr} we show the ratio of mean arrival time results for correlation-aware SARSOP policies to those using a correlation-unaware policy. We see that when used to directly optimize the POMDP using SARSOP, this approach typically yields moderate performance gains of 5--15\%. (In a few cases, the performance actually degrades; this is possible since the correlation-aware POMDP is more computationally complex and thus more difficult to solve with SARSOP.)  In the isotropic problem, the gains made are nevertheless insufficient to match the performance of SAI, which continues to perform better in this highly correlated case (not shown). We note in passing that SAI can be extended to be correlation-aware in a similar way and surprisingly the performance markedly degrades. Tests (not shown) suggest that this problem can be alleviated by increasing the infotactic optimization horizon to two or more time steps; work in this direction will be presented elsewhere.

\begin{figure*}
    \centering
    \includegraphics[width=\linewidth]{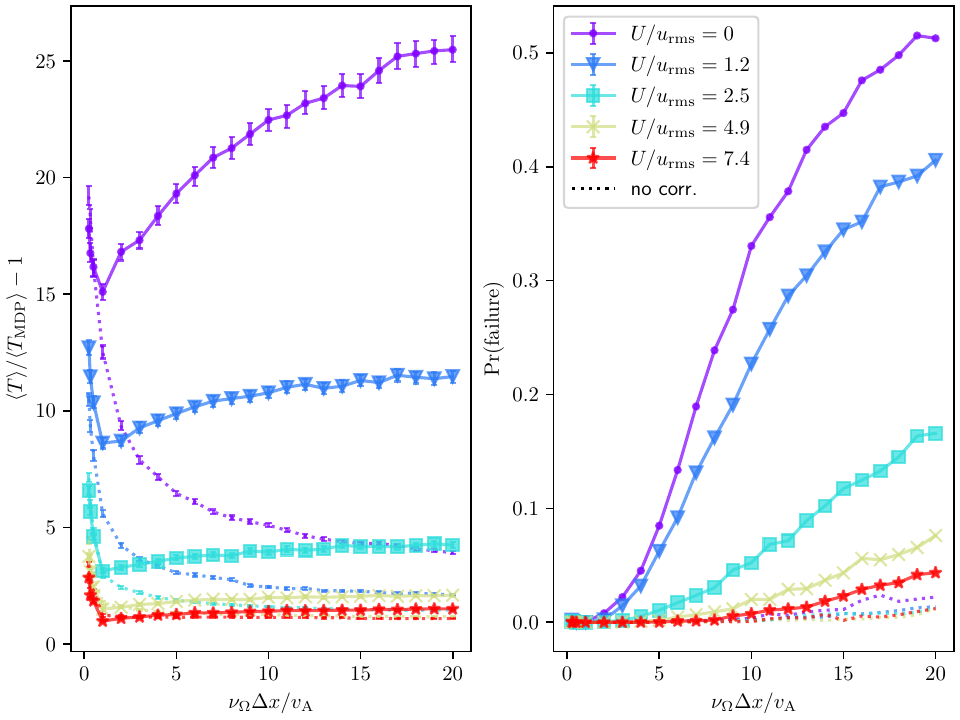}
    \caption{Left panel: mean excess arrival time for all wind speeds, as a function of the observation rate $\nu_\Omega,$ showing also trials performed in an uncorrelated setting where observations are drawn artificially from the empirical likelihood (dashed lines). Right panel: failure rates for the same sets of trials.}
    \label{fig:mult_obs_dummy}
\end{figure*}

\subsection{Parameter selection}
We have seen that with some exceptions, the presence of strong correlations tends to increase the typical time of arrival to the source. Besides simply reducing the time of arrival like $v_{A}^{-1},$ an increase to the agent's speed $v_A$ will also tend to decorrelate observations, which can only further help the agent. It therefore stands to reason that the agent should simply move as fast as possible, with the main limitations being mechanical.

The selection of $\nu_\Omega$ and $c_{\rm thr}$ at fixed $v_A$ are more interesting questions, which we now attempt to answer.

\subsubsection{Selection of $\nu_\Omega$}\label{sec:obs_rate}
 When the observation frequency $\nu_\Omega$ is very large, observations will not have time to decorrelate. This will tend to make each individual observation less useful, potentially degrading search performance. As a competing effect, a larger $\nu_\Omega$ means more observations and more information, improving performance. Which effect dominates? 
 
Under the first-order Markov model, we have from Eq.~\ref{eq:tstar2} (assuming positive correlations and replacing $\rho$ with an exponential) that the time to estimate the source is
\[T \simeq A \nu_{\Omega}^{-1} \frac{1+ \exp(-1/(\tau_c \nu_\Omega))}{1- \exp(-1/(\tau_c \nu_\Omega))} \]
for some $A.$ For large $\nu_\Omega,$ the arrival time thus should converge to a constant set by the correlation time $\tau_c:$ $T \simeq 2A\tau_c + O(\nu_\Omega^{-2}).$ 

We therefore expect diminishing returns on increasing the observation frequency far beyond a typical inverse correlation time. In practice, we argue that there is likely to be a finite optimum. There are at least two reasons why this may be the case. 

A first mechanism, which is perhaps specific to Bayesian policies, is an increase in the failure rate. Highly correlated observations made at a large rate in time can result in a posterior which, at early times, is strongly peaked in the wrong location (see the discussion in Sec.~\ref{sec:theory}). This can sometimes lead the agent to a region where observations are extremely rare and cause the agent to be trapped.

 We observe the latter effect in the following experiment: we fix the speed of the agent relative to the flow, interpolate the concentration data, and allow an agent to make $K\ge 1$ observations per action. [This is different from the experiments of Secs.~\ref{sec:unaware_exp}--\ref{sec:aware_exp} in that only the observation rate, and not the action rate, is being changed.] To be precise, the agent takes an action every $\Delta t = \tau_\eta$ (fixed) and makes an observation at each $\Delta t/K.$ To extend the data to small $\nu_\Omega,$ we also perform trials where the agent makes an observation only once every $K > 1$ actions. We show results for all wind speeds using the SARSOP policy, performing $10^4$ MC trials each; we have checked that SAI policies yield qualitatively similar results. The resulting arrival time performances are plotted in Fig.~\ref{fig:mult_obs_dummy}. We find that, indeed, there is an optimal observation frequency $\nu_\Omega\sim v_A/\Delta x,$ i.e., the rate at which actions are taken. In the right panel of the same figure, we show how the failure rate increases strongly past this optimum; it is this increase which is responsible for the increase in the (renormalized) mean arrival time. For comparison, we also show results for the uncorrelated problem; here, the performance simply improves monotonically and appears to converge to a small constant.

As a second mechanism for generating an optimal $\nu_\Omega,$ we suggest that making observations may incur a cost in a real agent (say, due to energy expenditure); minimizing a linear combination of the arrival time and this cost would fix the optimal observation frequency at a scale set by $\tau_c.$ In any case, we conclude that short-range correlations sharply limit any gains in performance from increasing the observation frequency beyond a typical inverse correlation time, potentially inducing a finite optimum observation frequency. The precise optimum will likely vary according to the details of the problem (in our setup, for example, the optimum was apparently not simply set by a correlation time).

\subsubsection{Selection of $c_{\rm thr}$}\label{sec:cthr}
\begin{figure*}
    \centering
    \includegraphics[width=\linewidth]{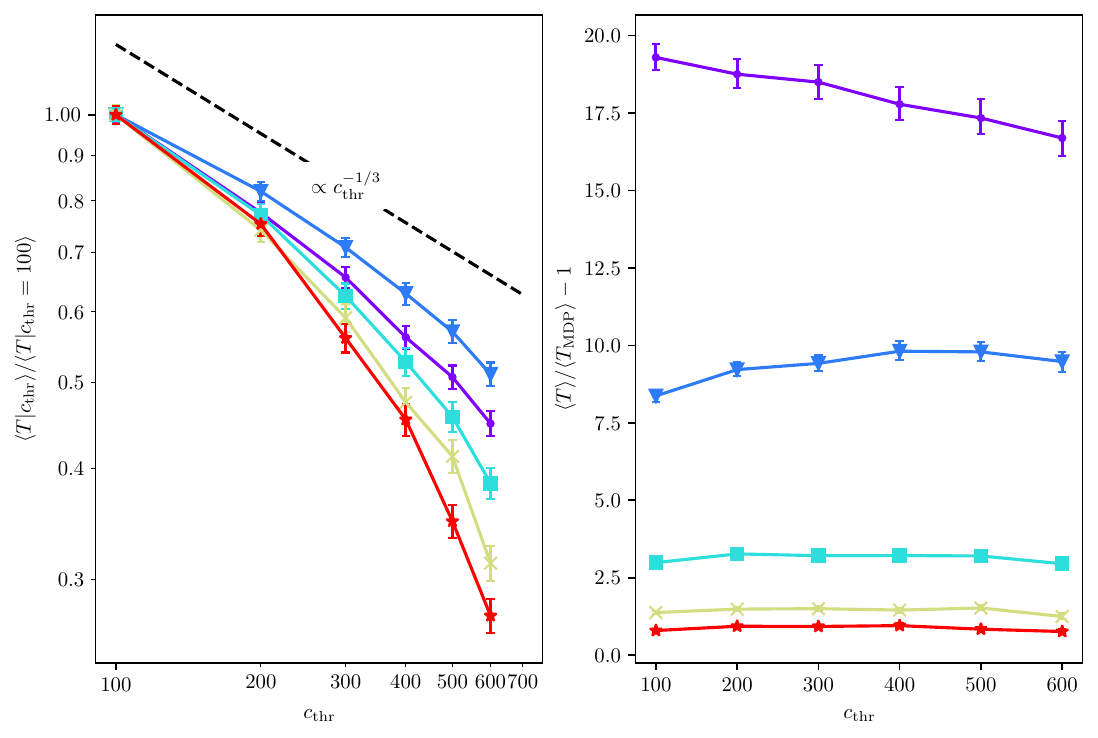}
    \caption{Left panel: mean (total) arrival time $\langle T \rangle$ under the SARSOP policy as a function of $c_{\rm thr},$ for each mean flow speed $U.$ Right panel: the same, but normalized by $\langle T_{\rm MDP} \rangle.$ Error bars indicate 95\% confidence intervals on the mean.}
    \label{fig:threshold_study}
\end{figure*}

\begin{figure}
    \centering
    \includegraphics[width=\linewidth]{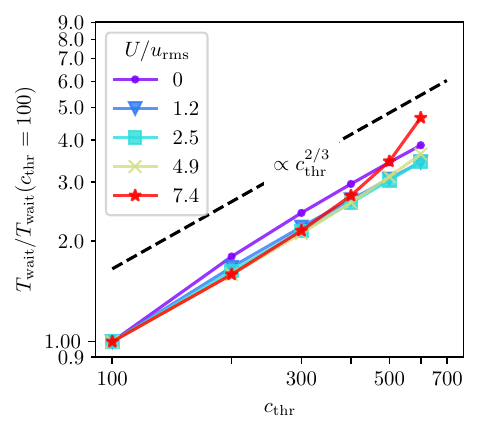}
    \caption{Estimate of the wait time $T_{\rm wait} \equiv \nu_\Omega^{-1} \langle p_1(\bm{x};c_{\rm thr}) \rangle^{-1},$ in units of $\nu_\Omega^{-1},$ as a function of $c_{\rm thr}.$}
    \label{fig:wait_times}
\end{figure}
Having established that correlations induce an optimum observation frequency, we proceed to study the proper choice of the threshold concentration $c_{\rm thr}.$ While we have previously established that correlations tend to increase in strength as $c_{\rm thr}$ decreases, we have checked that the dependence is not so strong: empirically, the (action-dependent) decorrelation rate grows slower than a power-law in $c_{\rm thr}.$ Instead, the main effect of $c_{\rm thr}$ on the arrival time to the source is relatively trivial: increasing $c_{\rm thr}$ concentrates the prior closer to the source, where a very large concentration event is much more likely to occur. This tends to decrease the arrival time $T_{\rm search}$, simply because the set of starting locations is drawn closer to the source. Nevertheless, increasing the threshold in this way comes at an opportunity cost: the agent must wait longer for the initial encounter. Therefore, it is interesting to ask what is the dependence on $c_{\rm thr}$ when we want to optimize a total time given by the sum of the two components
\[
T_{\rm total} = T_{\rm search} + T_{\rm wait},
\]
where the waiting time can be estimated as the rate of encounters, uniformly averaged over the arena:
\be
\label{eq:twait}
T_{\rm wait}^{-1} \simeq \frac{\nu_{\Omega}}{L^2} \int d \bm{x} \, p_1(\bm{x};c_{\rm thr})\ee
and $L$ is the size of the arena. $p_1(\bm{x};c_{\rm thr})$ is just the usual likelihood, with the dependence on $c_{\rm thr}$ made explicit. 

The scalings of $T_{\rm search}$ and $T_{\rm wait}$ with $c_{\rm thr}$ can be estimated as follows. Let the radius of a puff released from the source grow like $r\sim t^{1/\gamma}$ for some $\gamma\in(0,1].$ Then using $c\sim r^{-3},$ the concentration in the puff reaches a level $c_{\rm thr}$ at time $\tau\sim c_{\rm thr}^{-\gamma/3}$. Supposing the single Lagrangian particle dispersion to scale as $x \sim t^\alpha$ (with $\alpha\in (0,1]$), this corresponds to a distance of order $\ell_c \sim c_{\rm thr}^{-\gamma/(3\alpha)}.$ This sets a characteristic length-scale for the problem (however, it is not the only such scale: in the strong wind regime, the likelihood of an encounter decays in the downwind direction over a characteristic length $\propto c_{\rm thr}^{-\gamma/(\alpha(3-\alpha))}$ \cite{celani2014}). In particular, for weak wind, diffusion should dominate both single- and multi-particle dispersion, i.e.\ $\alpha=\gamma=2,$ so we expect $T_{\rm MDP}$ to scale like $c_{\rm thr}^{-1/3}.$ The dependence of $T_{\rm search}$ on $c_{\rm thr}$ should be dominated by the effect of $T_{\rm MDP}$ and obey the same scaling. Using a similar argument and simple dimensional analysis, we can also estimate $T_{\rm wait} \sim \nu_{\Omega}^{-1} \ell_c^{-2} \sim c_{\rm thr}^{2\gamma/(3\alpha)},$ yielding $c_{\rm thr}^{2/3}$ in the weak wind regime (or whenever $\gamma=\alpha$).

Altogether, we therefore roughly expect that
\[
T_{\rm total} \simeq f(\nu_{\Omega}) c_{\rm thr}^{-\gamma/(3\alpha)} + b L^2 \nu_{\Omega}^{-1} c_{\rm thr}^{2\gamma/(3\alpha)},   
\]
for some constant $b$ and some function $f$. In any case, for all choices of $\nu_{\Omega}$ there exists a unique optimal value $c^*_{\rm thr}$ which minimizes $T_{\rm total}.$ The optimum will scale with the arena size like $c^*_{\rm thr} \propto L^{-2\alpha/\gamma}.$

In Fig.~\ref{fig:threshold_study}, we show the empirical mean arrival time $\langle T \rangle$ under the optimized SARSOP policy as a function of $c_{\rm thr},$ for each mean flow speed $U$ (using the usual timestep of $\tau_{\eta}$). The results have been normalized by the value at $c_{\rm thr}=100,$ the smallest tested threshold. The result indeed agrees reasonably well with the diffusive estimate $\propto c_{\rm thr}^{-1/3}$ at small $U$ (at larger mean flow speed, ballistic transport becomes more important). We also show results for the excess arrival time, normalized by $\langle T_{\rm MDP} \rangle.$ After performing this rescaling, the dependence on $c_{\rm thr}$ is very weak, confirming that the decay in arrival time is almost entirely due to the decrease in $T_{\rm MDP}.$ In Fig.~\ref{fig:wait_times}, we compute the quantity $\nu_\Omega T_{\rm wait}$ using Eq.~\ref{eq:twait}. The agreement with the diffusive estimate, $\propto c_{\rm thr}^{2/3},$ is surprisingly good for all available wind speeds, at least over the range of thresholds we have tested. Importantly, we confirm that $T_{\rm search}$ is monotonically decreasing with $c_{\rm thr}$ and $T_{\rm wait}$ is monotonically increasing, guaranteeing the presence of an optimum.

\section{Discussion}\label{sec:discussion}
\appendix
Our results suggest that correlations in the flow can have a substantial impact on the search performance of a Bayesian agent, depending on the strength, sign, and range of the correlations. In general, one roughly expects positive correlations to be harmful and anticorrelations to be helpful, at least over short ranges.
\begin{figure*}
    \centering
    \includegraphics[width=0.9\linewidth]{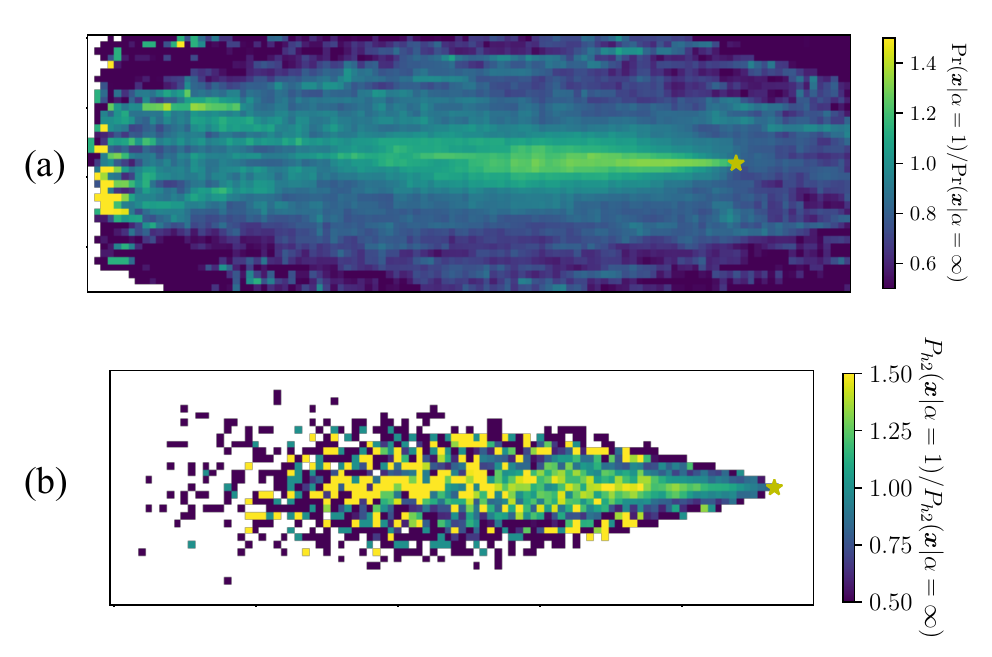}
    \caption{Upper panel (a): plot of the ratio ${\rm Pr}(\bm{x}|\alpha=1)/{\rm Pr}(\bm{x}|\alpha=\infty),$ where ${\rm Pr}(\bm{x})$ represents the fraction of time spent at $\bm{x}$ across an ensemble of $10^5$ searches. Lower panel (b): plot of the ratio $P_{h2}(\bm{x}|\alpha=1)/P_{h2}(\bm{x}|\alpha=\infty),$ where $P_{h2}(\bm{x})$ represents the probability of the second encounter (i.e., the first after time zero) occurring at point $\bm{x}.$ Both plots are for the case $U\simeq7.4 u_{\rm rms}.$ The source is indicated by a yellow star. }
    \label{fig:alpha-ratios}
\end{figure*}
One particularly interesting result is the reduction of arrival times in the presence of a sufficiently strong mean flow---even an agent \emph{unaware} of the correlations is able to passively exploit them, as found in Sec.~\ref{sec:unaware_exp}. This effect, which is absent if only short-range correlations are present, is probably linked to large-scale structures in the concentration field. To help understand this, in Fig.~\ref{fig:alpha-ratios}(a) we plot for $U\simeq 7.4u_{\rm rms}$ the ratio ${\rm Pr}(\bm{x}|\alpha=1)/{\rm Pr}(\bm{x}|\alpha=\infty),$ where ${\rm Pr}(\bm{x})$ represents the fraction of time spent at $\bm{x}$ across an ensemble of $10^5$ searches. When $\alpha=1,$ the agent is substantially less likely to spend time significantly far off the centerline or upwind of the source, where encounters are very rare and the search is especially difficult, relative to an agent in an uncorrelated environment. Instead the $\alpha=1$ agent spends more time around the centerline. This, in turn, is linked to the fact that the agent in the $\alpha=1$ environment is more likely to experience its second encounter (i.e., the first after time zero) near the centerline. We visualize this in Fig.~\ref{fig:alpha-ratios}(b) by plotting the ratio $P_{h2}(\bm{x}|\alpha=1)/P_{h2}(\bm{x}|\alpha=\infty).$ Here, $P_{h2}(\bm{x})$ represents the probability of the second encounter occurring at $\bm{x}.$ Indeed, the ratio is almost always larger than unity near the centerline. Therefore, it appears that large-scale structure can help orient the agent along the centerline, guiding the agent to the source faster.

On the other hand, strong positive correlations over short ranges can have a significant negative impact on the search when sweeping by the mean flow is not important, i.e., when the observation rate is high and/or $U$ is not much larger than $u_{\rm rms}$ (see Figs.~\ref{fig:windy_mreat_100},\ref{fig:isotropic_mreat_100},\ref{fig:mult_obs_dummy}).  Such correlations slow the convergence of the posterior to the ground truth, increasing the arrival time to the source. This effect can be mitigated if the agent has prior knowledge of the spatial dependence of the correlations through higher-order optimization, but only partially, as discussed in Sec.~\ref{sec:aware_exp}. 

We have seen that the presence of correlations also limits the utility of measuring the concentration field at a high rate in time and found evidence for an optimal rate of observation (Sec.~\ref{sec:obs_rate}). In Sec.~\ref{sec:cthr}, we argued additionally for the existence of an optimal threshold $c_{\rm thr}.$ In deriving this result, we assumed the threshold is fixed. While this is a reasonable assumption in robotics contexts, flying insects are not bound by such a constraint: fruitfly models have been observed to adapt their odor sensitivity dynamically to a local mean concentration \cite{nagel2015,cao2016,gorur2017}. However, noting that the optimum threshold decays as a power law in the arena size, one can imagine dynamically adapting the threshold as the volume covered by the posterior changes (typically, this will tend to decrease as the agent acquires more information). Qualitatively, this procedure would be similar to adaptive sensing in insects.


We studied the effect of correlations on the time to estimate the source position from the perspective of a one-step Markov model, assuming a stationary agent (Sec.~\ref{sec:theory}), and used it to develop a crude semi-empirical estimate for the arrival time (Eq.~\ref{eq:markov_est}). Improving this model and validating it against data is no small task: higher-order Markov models rapidly become more complicated analytically as well as prohibitively data-hungry for validation purposes. Moreover, the motion of the agent introduces complex effects due to the variation in $p_1$ over the trajectory and the coupling of correlations to the agent's motion; our approach of taking a spatial average over the trajectory is a severe and uncontrolled approximation. 

In future work, it will be important to explore more deeply the possibility of \emph{actively} exploiting correlations in the flow, which would likely require a model-free reinforcement learning agent with memory. The model-free approach is known to be difficult for the olfactory search problem because the cost signal is very sparse and a good policy requires substantial memory; it has nevertheless seen some recent study, for instance in Refs.~\cite{singh2023,verano2023,rando2024}. Additionally, we suggest that teams of multiple cooperating agents may be able to exploit correlations between simultaneous measurements at multiple points in space; multi-agent problems will be studied in forthcoming work.

Finally, a notable shortcoming of the present work is the choice to work in two dimensions, a necessary evil which dramatically reduces the computational complexity of the POMDP solution but does not take full advantage of our three-dimensional dataset. Extending the research to three dimensions is a target for future study. We also intend to exploit the presence of multiple sources of particles in our data; problems with multiple sources have seen some study \cite{masson2013,song2019} but likely never with realistic data. Problems with multiple species of passive scalar cues are another possible avenue for future research.

\section{Asymptotic analysis of the posterior}\label{app:theory}
Consider a stationary agent at position $x^*$ relative to the source, and let the observation sequence be $\Omega_i.$ Let the true likelihood at $x$ be $p_{\Omega' \Omega}(x) = p(\Omega_i|\Omega_{i-1},x) = {\rm Pr} (\Omega_i = \Omega'|\Omega_{i-1} = \Omega,x),$ i.e., the observations are first-order Markovian, and let the agent update its posterior using a model likelihood $\ell(\Omega' | \Omega,x),$ which for the moment we do not specify. Let us also introduce the notations $p_\Omega(x) = {\rm Pr}(\Omega_i = \Omega|x)$ for the true \emph{single-time} likelihood at $x$.

Now, consider the agent's posterior $b_N(x)$ after $N$ observations $\{\Omega_i \}_{i=1}^N$. We have
\be b_N(x) = \frac{\exp \left(-\sum_{i=1}^N \lambda_i(x)  \right) b_0(x)}{\sum_{x'} \exp \left(-\sum_{i=1}^N \lambda_i(x')  \right) b_0(x')}, \ee
where $\lambda_i(x) \equiv -\log \ell(\Omega_{i}|\Omega_{i-1},x).$

The object of interest is the Bayes factor
\begin{align*}
{\cal B}(x;x^*) &\equiv \frac{b_N(x)}{b_N(x^*)} \\ &= \frac{b_0(x)}{b_0(x^*)} \exp \left(\sum_{i=1}^N (\lambda_i(x^*) - \lambda_i(x))  \right),
\end{align*}
for when $x\ne x^*$ we want $\cal B$ to converge to zero as quickly as possible. 

Let us introduce a shorthand: the superscript $^*$ will indicate evaluation at $x^*$ and its absence will indicate evaluation at a test point $x$. By assumption, $\Omega_k$ is a first-order Markov chain and it can be shown \cite{jones2004} that as long as $\rho^*<1$ and $0<p_1^*<1$, the expression $\frac 1 N \sum_{i=1}^N \left(\lambda_i(x^*) - \lambda_i(x) \right)$ obeys a central limit theorem, yielding
\be \label{eq:clt}
\log {\cal B}(x;x^*) \to - N \mu + \sqrt{\Sigma N } \eta,
\ee
with $\eta\sim {\cal N}(0,1).$ Here, we have neglected the constant term coming from the prior and defined the expressions
\begin{widetext} \be \Sigma(x;x^*) \equiv \lim_{N\to \infty} \frac{1}{N} \sum_{i,j=1}^N \mathrm{Cov}(\lambda_i(x^*) - \lambda_i(x),\lambda_j(x^*) - \lambda_j(x))\ee \end{widetext}
and 
\be
\mu(x;x^*) \equiv \lim_{N\to \infty} \frac1N\sum_{i=1}^N {\rm E}(\lambda_i(x) - \lambda_i(x^*)). 
\ee
The averages should be understood as taken over possible observation histories. 

As an immediate consequence of Eq.~\ref{eq:clt}, the fluctuations from $\mu$ are insignificant when $N > \Sigma/\mu^2.$ We therefore define 
\[ T_{\rm converge}(x;x^*) = \frac{\Sigma}{\mu^2}\]
as the typical time for the posterior to converge in the sense of probability at the test point $x$.

We now specialize to the case where the agent is unaware of correlations but has the correct single-time likelihood, so that the model is $\ell(\Omega' | \Omega,x) = p_{\Omega'}(x).$ Since the $\Omega_i$ are ergodic, we have
\begin{align*}\label{eq:asymptotic_unaware}\mu = \lim_{N \to \infty} {\rm E}(\lambda^*_N-\lambda_N) &= \sum_\Omega p_\Omega^* \log \frac{p_\Omega^*}{p_\Omega} \\ &= D_{\rm KL}\left(p(\cdot| x^*) \parallel p( \cdot | x)\right)\end{align*}
where $D_{\rm KL}(p\parallel q)$ denotes the Kullback-Leibler (KL) divergence of $p$ from $q$, for distributions $p$ and $q$. This expression depends only on the single-time likelihoods, so the correlations have no effect on $\mu$, and their contribution to $T_{\rm converge}$ comes from $\Sigma$. 

 

When $N \gg T_{\rm converge}$, we have
\be
{\cal B}(x;x^*) \simeq \frac{b_0(x)}{b_0(x^*)}\exp \left(-N \mu  \right),
\ee
and since $\mu \ge 0$ (Gibbs' inequality), probability will tend to concentrate on the curve of equal likelihood that contains the true source location, i.e., the set $M = \{ x : \mu(x;x^*) = 0\} = \{ x : p_1(x) = p_1^*\}.$  $M$ will be a set of codimension one and thus have measure zero, and we have $x^* \in M,$ so the posterior is ``consistent'' in the jargon of Bayesian statistics. We refer the reader to Fig.~\ref{fig:stationary_agent} for a visualization of how the posterior collapse on the set $M.$

We therefore define a second characteristic time
\[ T_{\rm decay}(x;x^*) = \mu^{-1},\]
which is the time it takes for the posterior to decay at a test point $x\notin M,$ assuming it has converged in probability. We will eventually seek a time $T^* \ge \max\{T_{\rm converge},T_{\rm decay}, \}$ guaranteeing the convergence of the posterior onto the set $M.$

But first we must compute $\Sigma,$ which can be done by using the (left) transition matrix for $\Omega_k$: $\mathbf{T} = \begin{pmatrix} p_{00} & p_{01} \\ p_{10} & p_{11} \end{pmatrix}.$ One can diagonalize to compute
\be
\mathbf{T}^k =  \begin{pmatrix}p_0 +  p_1\rho^k& p_0 (1-\rho^k) \\ p_1 (1-\rho^k) & p_1+p_0 \rho^k \end{pmatrix}
\ee
for all $k\ge0$ (note that $\rho$ is an eigenvalue of $\mathbf{T}$, along with unity).

We have
\begin{widetext} \[ \Sigma = \lim_{N \to \infty} \left({\rm Var} (\lambda_N^*-\lambda_N)+ \frac2N\sum_{i=1}^N \sum_{j = i+1}^N \mathrm{Cov}(\lambda_i^*, \lambda_j) \right). \]\end{widetext}
Using ergodicity, the first term can be evaluated by averaging with respect to the invariant probability $p_i.$ The second term turns out to be a geometric series which can be computed directly, with the aid of $T^k.$ This leads to the following expression for $\Sigma,$ when the agent is unaware of correlations:
\be \Sigma = p_0^* p_1^* \log^2 \frac{p_0 p_1^*}{p_1 p_0^*}\left(1+\frac{2\rho^*}{1-\rho^*}\right). \ee
[Note that $\Sigma$ is independent of the prior, a consequence of ergodicity.] Denoting $\Sigma_0= p_0^* p_1^* \log^2 \frac{p_0 p_1^*}{p_1 p_0^*} $  the baseline in the absence of correlations, we have
\be \Sigma_{\rm unaware}-\Sigma_0 = \frac{2 \rho^*}{1-\rho^*} \Sigma_0.  \ee
This tells us that positive (negative) correlations will decrease (increase) the rate of convergence of the posterior.

As a side remark, note that the observations cannot be arbitrarily strongly anticorrelated, due to the constraint
\be \rho \ge -\min \left( \frac{p_1}{p_0},\frac{p_0}{p_1} \right),
\ee
which follows from the requirement $0 \le p_{10},p_{11} \le 1$ along with the law of total probability
\be
p_1 = p_{10} p_0 + p_{11} p_1.
\ee
This constraint means that if $\rho<0,$ $|\rho|$ must be small if $p_1$ is close to 0 or 1. It is also worth mentioning the special case $\rho = -1,$ which is only possible when $p_1 =1/2,$ since $\Omega_i$ must simply alternate between 0 and 1 at every timestep. 

We now derive $T^*.$ After a time $T,$ the agent's posterior is converged in probability and sufficiently decayed at $x$ only if $T> T_{\rm converge}$ and $T> T_{\rm decay}.$ We seek a single, global upper bound on the requisite $T.$ The main challenge is that $T_{\rm converge}$ blows up near $M,$ i.e., when the likelihood is very precisely estimated. This motivates imposing a tolerance $\epsilon>0$ on the precision of the estimated likelihood: we put $|p_1^*-p_1| \ge \epsilon$ for some $\epsilon>0.$ It is then possible to establish the (sharp) inequalities
\begin{align}
\label{eq:ineq1} T_{\rm converge} & \le \frac{1}{\epsilon^2} \frac{1+\rho^*}{1-\rho^*}, \\
 \label{eq:ineq2}   T_{\rm decay} &\le \frac{1}{2\epsilon^2},
\end{align}

The inequality (\ref{eq:ineq1}) can be proven by, for example, defining 
\begin{widetext}\[ f(y) = (x - y) \sqrt{x(1-x)} \log\frac{x(1-y)}{y(1-x)} - x\log \frac{x}{y} - (1-x) \log \frac{1-x}{1-y}, \]\end{widetext}for $x,y\in(0,1)$ and differentiating to show that $f$ has a unique maximum at $y=x$, namely zero. (The form of $f(y)$ comes from demanding $\sqrt{\Sigma} - \mu \ge 0.$) Inequality (\ref{eq:ineq2}) is just a special case of Pinsker's inequality. Together, these inequalities establish Eq.~\ref{eq:tstar2}.
As a final remark, the asymptotic analysis for a moving agent is possible, at least in the limit of continuous observation, but considerably more involved and must be deferred to future work.

\acknowledgments{We thank Aurore Loisy, Massimo Cencini, Chiara Calascibetta, Lorenzo Piro, Mauro Sbragaglia, and Michele Buzzicotti for fruitful discussions. This work received funding from the European Union's H2020 Program under grant agreement No.\ 882340. We also acknowledge financial support under the National Recovery and Resilience Plan (NRRP), Mission 4, Component 2, Investment 1.1, Call for tender No. 104 published on 2.2.2022 by the Italian Ministry of University and Research (MUR), funded by the European Union – NextGenerationEU– Project Title Equations informed and data-driven approaches for collective optimal search in complex flows (CO-SEARCH), Contract 202249Z89M. - CUP B53D23003920006 and E53D23001610006. The data that support the findings of this article are openly available \cite{turbodor,code}, embargo periods may apply.}

\end{document}